\documentclass[12pt,a4paper]{article}

\usepackage[utf8]{inputenc}
\usepackage[T1]{fontenc}
\usepackage[english]{babel}

\usepackage{amsmath,amssymb,amsthm}
\usepackage{mathtools}
\usepackage{dsfont}
\usepackage{mathrsfs}

\usepackage{geometry}
\usepackage{microtype}
\usepackage{setspace}

\usepackage{tikz}
\usetikzlibrary{patterns, arrows.meta, decorations.pathmorphing, calc}

\usepackage{graphicx}
\usepackage{booktabs}

\usepackage{hyperref}
\hypersetup{
  colorlinks = true,
  linkcolor  = blue,
  citecolor  = blue,
  urlcolor   = blue
}

\usepackage[
  backend    = biber,
  style      = phys,
  sorting    = none,
  maxnames   = 5,
  doi        = true,
  url        = false
]{biblatex}
\usepackage{fancyhdr}
\fancypagestyle{plain}{%
  \fancyhf{}%
  \fancyhead[L]{\scshape A Preprint}%
  \fancyhead[R]{\today}%
  \fancyfoot[C]{\thepage}%
}

\newcommand{\R}{\mathbb{R}}
\newcommand{\diff}{\mathrm{d}}
\newcommand{\bn}{\mathbf{n}}
\newcommand{\br}{\mathbf{r}}
\newcommand{\bR}{\mathbf{R}}
\newcommand{\bu}{\mathbf{u}}
\newcommand{\bE}{\mathbf{E}}
\newcommand{\bH}{\mathbf{H}}
\newcommand{\bD}{\mathbf{D}}
\newcommand{\bP}{\mathbf{P}}
\newcommand{\dOmega}{\partial\Omega}
\newcommand{\Vmoy}{\mathscr{V}(\dOmega)}
\newcommand{\Rp}{R_\perp}

\newcommand{\chis}{\chi^s}
\newcommand{\chispar}{\chi^s_\parallel}
\newcommand{\chisper}{\chi^s_\perp}
\newcommand{\jump}[1]{\left[\!\left[#1\right]\!\right]}
\newcommand{\avg}[1]{\left\langle #1\right\rangle}

\newtheorem{remark}{Remark}

\begin{document}

\title{Nonlocal Optical Response and Surface Susceptibilities:\\
  A Systematic Derivation via Spatial Moment Expansion}

\author{F. Zolla}

\date{\today}

\maketitle

\begin{abstract}
We present a systematic theory connecting the nonlocal response kernel of
a homogeneous medium to its effective surface susceptibilities for an
arbitrary curved interface. Starting from the most general tensorial
nonlocal constitutive relation and combining a spatial moment expansion
with a distributional thin-layer limit, we show that the full complexity
of the interfacial response condenses, at leading order, into a single scalar:
the surface susceptibility $\chis$, equal for the tangential and normal
components of the electric field.
These quantities provide a constructive generalization of the Feibelman
$d$-parameters to interfaces of arbitrary curvature, and the curvature
corrections, proportional to the geometric invariants $H$ (mean
curvature) and $K$ (Gaussian curvature), are derived explicitly. The
formalism is illustrated on a comprehensive set of analytically tractable
cases (planar, spherical, cylindrical, and ellipsoidal interfaces) for
several kernel choices (Gaussian, Yukawa, tensorial Lorentz). Generalized
Maxwell boundary conditions are established and compared with the
classical Fresnel results.
\end{abstract}

\section{Introduction}
The interaction between light and matter is one of the oldest subjects in
physics. Despite the abundance of work devoted to this question over more than
a century, it would be an overstatement to claim that the field is fully
understood today. In many practical situations, theoretical descriptions still
rely on partially empirical models whose microscopic justification remains
incomplete.

One of the deep reasons for this state of affairs lies in the coexistence of
several vastly different spatial scales in optical phenomena. In a typical
problem of electromagnetism applied to optics, one can distinguish at least
three characteristic scales.
\begin{itemize}
    \item The first is the atomic scale, of the order of the angstr\"om. At
    this scale, electromagnetic fields vary over distances comparable to the
    dimensions of electronic orbitals. Quantum mechanics reigns supreme.

    \item The second is the scale of optical wavelengths, typically of the
    order of a micrometre in the visible range.

    \item The third is the scale of the macroscopic objects being illuminated,
    which can reach several millimetres or centimetres.
\end{itemize}

A complete description of the light--matter interaction should in principle
allow these three levels of description to be connected. In particular, the
passage from the atomic scale to the optical scale should be obtained through
a procedure of statistical and quantum averaging, making it possible to define
the macroscopic quantities used by the optical physicist, such as the
permittivity or the susceptibility of the medium~\cite{BornWolf1999,LandauLifshitz1984}.

In practice, this programme remains largely unfulfilled for the systems of
interest to modern optics. \textit{Ab initio} approaches exist, but they remain
difficult to apply in realistic situations involving surfaces, interfaces, or
complex structures~\cite{Onida2002}.

One is therefore led to adopt a different approach, often described as
phenomenological, which consists in postulating certain constitutive relations
between the macroscopic electromagnetic quantities. More precisely, one seeks
to establish relations connecting the electric and magnetic fields
$(\mathbf{E}, \mathbf{B})$ to the displacement and induction fields
$(\mathbf{D}, \mathbf{H})$.

These relations must satisfy several fundamental requirements. In particular,
they must:
\begin{itemize}
    \item be independent of the choice of coordinate system;
    \item respect the symmetries of the physical system under consideration;
    \item be compatible with the general principles of physics, in particular
    the causality principle.
\end{itemize}

In this framework, the electric polarisation $\mathbf{P}$ plays a central
role, since it connects the fields $\mathbf{E}$ and $\mathbf{D}$ through the
relation
\[
    \mathbf{D} = \varepsilon_0 \mathbf{E} + \mathbf{P}.
\]
The main difficulty then consists in determining the functional relation
between $\mathbf{P}$ and $\mathbf{E}$.
The optical response of material media is traditionally formulated within
the local framework, in which the polarization vector at a point $\br$ depends
only on the electric field at that same point~\cite{BornWolf1999,LandauLifshitz1984}.
This approximation, which is remarkably successful for describing wave
propagation in macroscopic homogeneous media, breaks down as soon as the
relevant length scales become comparable to the microscopic characteristic
lengths of the material, typically a few \r{A}ngstr\"{o}ms to a few nanometres,
for metals and semiconductors.

\textit{Spatial non-locality} is a considerably more subtle issue. In this case, the
polarisation vector at a point in space depends on the electric field in a spatial
neighbourhood.
Unlike the temporal case, there is no principle of spatial causality that
could constrain the form of this relation. Furthermore, the three-dimensional
structure of space allows for a wide variety of geometric behaviours that have
no counterpart in the temporal problem.

%

\emph{Spatial dispersion}  which translates the dependence of the dielectric
susceptibility not only on frequency $\omega$ but also on the wavevector
$\mathbf{k}$ was introduced theoretically by Pekar \cite{Pekar1958}
and developed systematically by Agranovich and Ginzburg
\cite{Agranovich1984}. It expresses the fact that the polarization at
$\br$ results from a weighted integral of the field over a neighbourhood
of $\br$, whose spatial extent is set by the range of the response kernel.
This nonlocal framework has proved indispensable for describing excitonic
polaritons in semiconductors \cite{Hopfield1958,Halevi1992}, and more
recently, nonlocal effects in plasmonic nanostructures
\cite{Mortensen2014,Raza2015,Ciraci2012}.

The rapid growth of nanophotonics has renewed interest in the description
of interfaces in this nonlocal context. Even in a centrosymmetric medium,
the symmetry breaking induced by a surface generates a response
concentrated in an atomic-scale transition layer, qualitatively different
from the bulk. Feibelman \cite{Feibelman1982} showed that this surface
response can be characterized, at leading order, by two scalar parameters
$d_\perp$ and $d_\parallel$, the so called \emph{Feibelman $d$-parameters},
encoding the response of the transition layer to the normal and tangential
components of the electric field. These parameters have since been
computed \textit{ab initio} for a wide range of metal surfaces
\cite{Liebsch1997,Persson1983}, measured experimentally via ellipsometry
and differential reflectance spectroscopy \cite{YangNature2019,Bosman2022},
and incorporated into extended Maxwell boundary conditions for various
geometries \cite{Christensen2017,Babaze2023}.

Despite this progress, the connection between the bulk nonlocal response
kernel and the effective surface susceptibilities, and more generally
the generalized boundary conditions for an arbitrarily curved interface, has remained elusive. Most existing approaches rely either on
phenomenological models such as the nonlocal hydrodynamic model
\cite{Halevi1992,Mortensen2014,Toscano2015}, or on material-specific
microscopic calculations \cite{Liebsch1997,ApellPickard1985}, without
providing a constructive link between the bulk kernel and the surface
response for an arbitrary geometry. In particular, the influence of
interface curvature which is central for understanding the optical properties
of nanoparticles \cite{Ciraci2012,Christensen2017} and the nonlocal Mie
corrections  has not been derived systematically from the bulk kernel
alone.

The present paper fills this gap by providing a systematic and rigorous
derivation of the linear surface susceptibilities from the bulk nonlocal
response kernel. Our approach rests on two ingredients:

\begin{enumerate}
  \item \textbf{A spatial moment expansion} of the tensorial kernel
    $\Delta^{(1)}_{ij}(\br,\br')$, which extends the classical Taylor
    expansion of the field to the three-dimensional vector case and
    naturally decomposes the response into a bulk contribution and a
    boundary contribution;

  \item \textbf{A distributional thin-layer limit}, which converts
    functions concentrated near the interface into singular distributions
    supported on $\dOmega$, i.e. Dirac deltas and their normal derivatives,
whose coefficients are moment integrals of the kernel over the
    exterior half-space.
\end{enumerate}

This framework yields, for any regular surface geometry: (i) the surface
susceptibility $\chis$ at leading order, providing a constructive
generalization of the Feibelman parameters; (ii) curvature
corrections proportional to $H$ and $K$; and (iii) generalized Maxwell
boundary conditions in explicit form, relating field jumps to kernel
moments. All microscopic complexity is thereby condensed into a finite set
of effective parameters, measurable experimentally without knowledge of
the underlying microscopic model.

The present paper restricts attention to the case where the nonlocal kernel
is \emph{centrosymmetric}, i.e., invariant under parity
$\bR \mapsto -\bR$.  This symmetry, which holds for all standard
plasmonic metals and dielectrics, forces the rank-one moment to vanish
identically and makes the surface susceptibility $\chis$ the
leading-order descriptor of the interfacial response.
When the kernel is \emph{not} centrosymmetric, as occurs, for instance,
in a medium composed of identically handed chiral inclusions , a
rank-three moment survives, governed by a single pseudoscalar parameter
$\beta$ that encodes the chirality of the microscopic response.  This
complementary regime, which leads to optical activity and the
Drude--Born--Fedorov constitutive relation, is treated in the companion
paper~\cite{ArticleChiral}.

The paper is organized as follows. Section~\ref{sec:1D} treats the
one-dimensional scalar case: with trivial geometry, all distributional
calculations can be carried out explicitly, and the essential mechanisms
i.e.  bulk/boundary decomposition, thin-layer limit, expansion in powers of
the kernel range emerge without tensorial complications, together with
a set of explicit numerical examples. Section~\ref{sec:3D} extends the
formalism to the three-dimensional tensorial case, treating successively
the planar interface (reference case), the sphere, the cylinder and the
ellipsoid, with fully explicit calculations for Gaussian, Yukawa and
tensorial Lorentz kernels. Section~\ref{sec:CL} derives the generalized
Maxwell boundary conditions and establishes their connection with the
Feibelman $d$-parameters. Section~\ref{sec:conclu} concludes and outlines
the extension to the nonlinear case, which is the subject of a companion
paper \cite{ArticleNL}.



\section{The one-dimensional scalar case}
\label{sec:1D}

\subsection{Physical motivation and setup}

Before tackling the full three-dimensional tensorial problem, we develop
the essential ideas in the simplest possible setting: a one-dimensional
scalar model. This is not merely a pedagogical device. It corresponds
precisely to the physically relevant situation of TE-polarized light
incident on a planar interface between vacuum and a nonlocal medium. In
this geometry the electric field has a single nonzero component parallel
to the interface, and the constitutive relation reduces to a scalar
integral operator.

We consider two half-spaces: vacuum for $z < 0$ and a nonlocal medium
occupying $\Omega = \{z > 0\}$, with the interface at $z = 0$.
The nonlocal constitutive relation takes the form
\begin{equation}\label{eq:constit1D}
  P(z) = \varepsilon_0 \int_{\R} \Delta(z,z')\, E(z')\, \diff z'.
\end{equation}

\subsection{Spatial moment expansion}

We expand the field $E(z')$ in a Taylor series about the observation
point $z$:
\begin{equation}\label{eq:Taylor1D}
  E(z') \approx E(z) + (z'-z)\, E'(z)
  + \tfrac{1}{2}(z'-z)^2\, E''(z) + \cdots
\end{equation}
Inserting this into~\eqref{eq:constit1D} yields the
\emph{spatial moment expansion}:
\begin{equation}\label{eq:P_moments1D}
  P(z) \approx \varepsilon_0
  \bigl[
    \chi_0(z)\, E(z)
    + \chi_1(z)\, E'(z)
    + \chi_2(z)\, E''(z)
    + \cdots
  \bigr],
\end{equation}
where the \emph{spatial moments} of the kernel are defined by
\begin{equation}\label{eq:chi_j_def}
  \chi_j(z)
  := \frac{1}{j!}\int_{\R}(z'-z)^j\,\Delta(z,z')\,\diff z',
  \qquad j = 0,1,2,\ldots
\end{equation}

\begin{remark}[Validity of the expansion]
  For a plane-wave component $E(z) = e^{ikz}$, the expansion
  \eqref{eq:Taylor1D} is accurate provided $k|z'-z|\ll 1$, i.e., the
  nonlocality range $\ell$ satisfies $\ell \ll \lambda$.  This condition
  holds in all optical situations of interest, since $\ell$ is of the
  order of a few \r{A}ngstr\"{o}ms while optical wavelengths are at least
  several nanometres.
\end{remark}

\subsection{Hypotheses on the kernel}

We impose the following conditions on $\Delta(z,z')$, in increasing order
of specificity.

\begin{enumerate}

\item \textbf{Rapid decay.}
  For fixed $z$, the function $\Delta(z,z')$ decays rapidly:
  all moments with respect to $(z'-z)$ exist.

\item \textbf{Support in $\Omega\times\Omega$.}
  The polarization vanishes in vacuum, and the vacuum region does not
  drive the nonlocal response:
  \begin{equation}
    \Delta(z,z') = H(z)\,H(z')\,\Delta(z,z'),
  \end{equation}
  where $H$ denotes the Heaviside function.

\item \textbf{Bulk homogeneity.}
  Far from the interface the medium is translationally invariant:
  \begin{equation}\label{eq:homo1D}
    \Delta(z,z') = H(z)\,H(z')\,\tilde\Delta(z-z').
  \end{equation}
  This guarantees that the medium becomes homogeneous away from $z=0$.

\item \textbf{Centro-symmetry.}
  The bulk kernel $\tilde\Delta$ is even:
  $\tilde\Delta(-Z) = \tilde\Delta(Z)$.

\end{enumerate}

Under these hypotheses and the change of variable $Z = z'-z$:
\begin{equation}\label{eq:chi_j_explicit}
  \chi_j(z) = \frac{H(z)}{j!}
  \int_{-z}^{+\infty} Z^j\,\tilde\Delta(Z)\,\diff Z.
\end{equation}

\begin{remark}
  The constraint $z'>0$ translates into
  $Z = z'-z > -z$, hence the lower limit $-z$ in~\eqref{eq:chi_j_explicit}.
  It is this lower limit, dependent on the distance to the interface, that
  carries all the surface physics.
\end{remark}

\subsection{Bulk and boundary decomposition}

We split the integration range in~\eqref{eq:chi_j_explicit} by writing
$\int_{-z}^{+\infty} = \int_{\R} - \int_{-\infty}^{-z}$:
\begin{equation}\label{eq:decomp1D}
  \chi_j(z)
  =
  \chi_j^{\Omega}\,H(z)
  \;-\;
  \chi_j^{\Vmoy}(z),
\end{equation}
where the \emph{bulk moment} is the constant
\begin{equation}\label{eq:chi_j_bulk}
  \chi_j^{\Omega}
  := \frac{1}{j!}\int_{\R} Z^j\,\tilde\Delta(Z)\,\diff Z,
\end{equation}
and the \emph{boundary term} is
\begin{equation}\label{eq:chi_j_bdy}
  \chi_j^{\Vmoy}(z)
  := \frac{H(z)}{j!}\int_{-\infty}^{-z} Z^j\,\tilde\Delta(Z)\,\diff Z.
\end{equation}
\begin{remark}
This dichotomy is crucial: it is clear that $\chi_j^{\Vmoy}(z)$ tends to zero
as $z$ tends to infinity (i.e. towards the non-local material at play), and
that consequently this susceptibility is attached to the boundary, hence the
presence of the superscript $\Vmoy$ (vicinity of $\partial \Omega$) , while
the other part $\chi_j^{\Omega}$ can only be attached to the ``bulk'' itself.
\end{remark}
\paragraph{Vanishing of odd bulk moments.}
By centro-symmetry, the integrand $Z^j\tilde\Delta(Z)$ is odd for odd $j$,
so $\chi_j^{\Omega}=0$ for $j$ odd. In particular, $\chi_1^{\Omega}=0$,
and the leading nonlocal correction to the bulk polarization is of second
order in the field gradient:
\begin{equation}\label{eq:P_bulk1D}
  P^{\Omega}(z)
  = \varepsilon_0
  \bigl(
    \chi_0^{\Omega}\,E(z) + \chi_2^{\Omega}\,E''(z) + \cdots
  \bigr),
\end{equation}
with
\begin{equation}
  \chi_0^{\Omega} = \int_{\R}\tilde\Delta(Z)\,\diff Z, \qquad
  \chi_2^{\Omega} = \frac{1}{2}\int_{\R} Z^2\,\tilde\Delta(Z)\,\diff Z.
\end{equation}
Equivalently, in the language of distributional convolutions:
\begin{equation}
  P^{\Omega}(z)
  = \varepsilon_0
  \bigl(
    \chi_0^{\Omega}\,\delta + \chi_2^{\Omega}\,\delta''
  \bigr)\star E\,(z).
\end{equation}

\paragraph{Localization of boundary terms.}
The boundary term $\chi_j^{\Vmoy}(z)$ is concentrated in a layer of
thickness $\sim\ell$ near $z=0$, and vanishes as $z\to+\infty$:
\begin{equation}
  \lim_{z\to 0^+}\chi_j^{\Vmoy}(z) = \chi_j^{\Omega}
  \;\text{(for even }j\text{)},
  \qquad
  \lim_{z\to+\infty}\chi_j^{\Vmoy}(z) = 0.
\end{equation}
These boundary terms are the fundamental objects that, after a
distributional thin-layer limit, give rise to the surface susceptibilities.

\subsection{Explicit example: the exponential kernel}
\label{sec:expo}

To fix ideas, we carry out all computations for the Yukawa-type kernel:
\begin{equation}\label{eq:expo_kernel}
  \tilde\Delta(Z) = \frac{A}{\ell}\,e^{-|Z|/\ell},
\end{equation}
where $\ell > 0$ is the nonlocality length and $A$ is a dimensionless
amplitude. This kernel is even, integrable, and admits finite moments of
all orders; $\tilde\Delta$ has dimension of an inverse length.

\subsubsection{Bulk moments}

\begin{align}
  \chi_0^{\Omega}
  &= \frac{A}{\ell}\int_{\R} e^{-|Z|/\ell}\,\diff Z = 2A,
  \label{eq:chi0_bulk_expo}\\[4pt]
  \chi_1^{\Omega} &= 0 \qquad\text{(by parity)},
  \label{eq:chi1_bulk_expo}\\[4pt]
  \chi_2^{\Omega}
  &= \frac{A}{2\ell}\int_{\R} Z^2\,e^{-|Z|/\ell}\,\diff Z = A\ell^2.
  \label{eq:chi2_bulk_expo}
\end{align}

\subsubsection{Boundary terms}

\begin{figure}[h!]
\begin{center}
\includegraphics[width=0.85\linewidth]{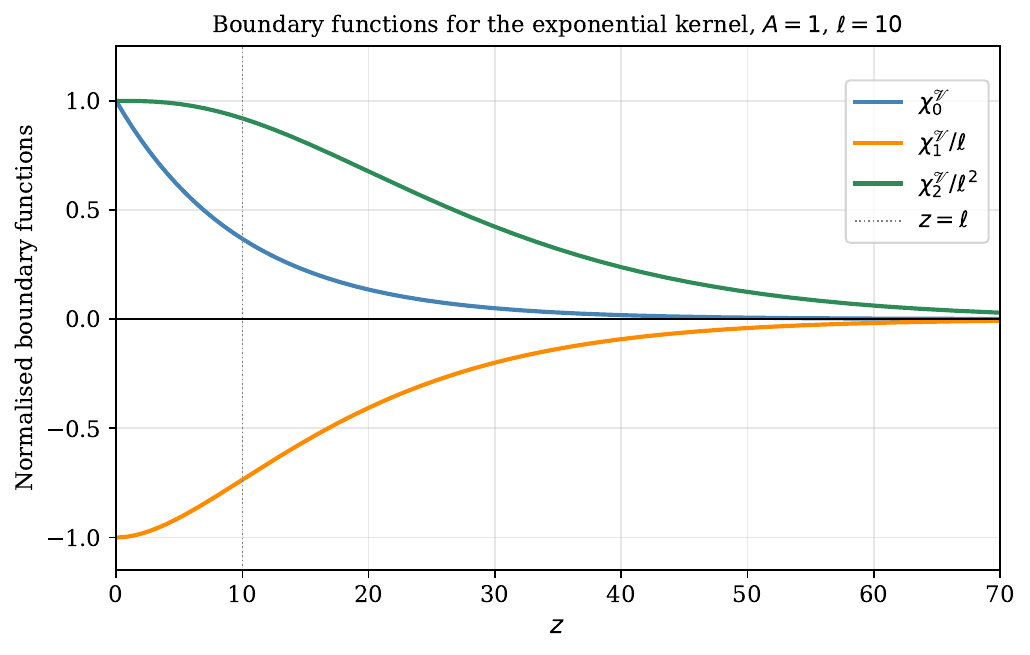}
\caption{Boundary functions $\chi_0^{\Vmoy}$,
         $\chi_1^{\Vmoy}/\ell$, and $\chi_2^{\Vmoy}/\ell^2$
         for the exponential kernel~\eqref{eq:expo_kernel}
         with $A=1$ and $\ell=10$.  Each function starts
         at $\pm A$ at $z=0$ and decays to zero on the
         scale $\ell$ (dotted vertical line).}
\label{fig:chi_j}
\end{center}
\end{figure}

By using the equality (\ref{eq:chi_j_bdy}), we obtain~:
\begin{align}
  \chi_0^{\Vmoy}(z)
  &= A\,e^{-z/\ell}\,H(z),
  \label{eq:chi0_bdy_expo}\\[4pt]
  \chi_1^{\Vmoy}(z)
  &= -A\ell\,e^{-z/\ell}\!\left(1+\frac{z}{\ell}\right)H(z),
  \label{eq:chi1_bdy_expo}\\[4pt]
  \chi_2^{\Vmoy}(z)
  &= A\ell^2\,e^{-z/\ell}\!\left(1+\frac{z}{\ell}
    +\frac{1}{2}\!\left(\frac{z}{\ell}\right)^{\!2}\right)H(z).
  \label{eq:chi2_bdy_expo}
\end{align}
These functions are plotted in Fig.~\ref{fig:chi_j} for $A=1$, $\ell=10$.

\subsection{Distributional thin-layer limit}
\label{sec:1D_distrib}

The boundary terms~\eqref{eq:chi_j_bdy} are functions localized in a
layer of thickness $\ell$ near $z=0$.  Since $\ell\ll\lambda$ in all
optical situations of interest, it is natural to collapse this layer onto
a mathematical surface and represent the boundary contribution as a
distribution supported at $z=0$.

\subsubsection{General principle}

We recall the standard result in distribution theory: if
$f\in L^1(\R)$, then
\begin{equation}\label{eq:dirac_limit}
  \frac{1}{\eta}\,f\!\left(\frac{z}{\eta}\right)
  \;\xrightarrow[\eta\to 0]{\mathcal{D}'}\;
  \left(\int_{\R} f(u)\,\diff u\right)\delta(z).
\end{equation}
We apply this to the rescaled boundary term
$\chi_\eta(z) := \frac{1}{\eta}\,\chi_0^{\Vmoy}(z/\eta)$.

\subsubsection{Distributional convergence}

Testing against a smooth compactly supported function $\varphi$:
\begin{equation}
  \int_{\R}\chi_\eta(z)\,\varphi(z)\,\diff z
  = \frac{1}{\eta}\int_0^{+\infty}
    F\!\left(\frac{z}{\eta}\right)\varphi(z)\,\diff z
  = \int_0^{+\infty} F(u)\,\varphi(\eta u)\,\diff u,
\end{equation}
where we set $u = z/\eta$ and introduced
\begin{equation}\label{eq:F_def}
  F(u) := \int_{-\infty}^{-u}\tilde\Delta(Z)\,\diff Z \;\geq 0,
  \qquad u\geq 0.
\end{equation}
By dominated convergence, as $\eta\to 0$:
\begin{equation}\label{eq:distrib_chi0}
  \frac{1}{\eta}\,\chi_0^{\Vmoy}\!\left(\frac{z}{\eta}\right)
  \;\xrightarrow[\eta\to 0]{\mathcal{D}'}\;
  A_0\,\delta(z),
\end{equation}
with
\begin{equation}\label{eq:A0_integral}
  A_0 = \int_0^{+\infty} F(u)\,\diff u.
\end{equation}

\subsubsection{Expression in terms of $\tilde\Delta$}

Inverting the order of integration (domain: $Z\leq -u$, $u\geq 0$
$\Leftrightarrow$ $Z\leq 0$, $0\leq u\leq -Z$):
\begin{equation}\label{eq:A0_fubini}
 \int_0^\infty F(u)\,\diff u
= \int_{Z=-\infty}^0 \left(\int_{u=0}^{-Z} \diff u \right) \tilde{\Delta}(Z)\, \diff Z =
-\int_{-\infty}^0 Z \, \tilde{\Delta}(Z)\,\diff Z=
\int^{+\infty}_0 Z\, \tilde{\Delta}(-Z)\,\diff Z,
\end{equation}
where the last equality uses centro-symmetry.  Hence the
\emph{leading surface coefficient} is the first moment of $\tilde\Delta$:
\begin{equation}\label{eq:A0_final}
A_0 = \int_0^{+\infty} Z\,\tilde\Delta(Z)\,\diff Z.
\end{equation}
In short
\begin{equation}
\chi_0^{\Vmoy}\left( \frac{z}{\eta}\right) = \eta \, A_0 \, \delta(z) + o(\eta) \, ,
\end{equation}
where $A_0$ is given by (\ref{eq:A0_final}). The singular part is the denoted $\chi_0^{\partial \Omega}(z) :=  A_0 \, \delta(z)$.

\subsubsection{Verification for the exponential kernel}

For $\tilde\Delta(Z) = (A/\ell)\,e^{-|Z|/\ell}$:
\[
  A_0 = \frac{A}{\ell}\int_0^{+\infty} Z\,e^{-Z/\ell}\,\diff Z
      = \frac{A}{\ell} \times \ell^2 = A \, \ell .
\]
One checks directly: $\chi_0^{\Vmoy}(z)=Ae^{-z/\ell}H(z)$, so
$\frac{1}{\eta}\chi_0^{\Vmoy}(z/\eta) = \frac{A}{\eta}\,e^{-z/(\eta \ell)}H(z)$,
which is a classical approximation of $A \, \ell \, \delta(z)$. \qed

\subsection{Full distributional expansion}
\label{sec:1D_full}

We now extend the leading-order result~\eqref{eq:distrib_chi0} to a
complete asymptotic expansion in powers of the layer thickness.

\subsubsection{Expansion theorem}

Assuming all moments of $\tilde\Delta$ are finite, the distributional
expansion of $\chi_\eta(z) = \frac{1}{\eta}\,\chi_0^{\Vmoy}(z/\eta)$
in the limit $\eta\to 0$ reads:
\begin{equation}\label{eq:full_expansion}
  \chi_\eta(z)
  = A_0\,\delta(z)
  + \eta\, A_1\,\delta'(z)
  + \eta^2\, A_2\,\delta''(z)
  + o(\eta^2)
\end{equation}
with
\begin{equation}\label{eq:Ak_general}
  A_k
  = \frac{(-1)^k}{k!}\int_0^{+\infty} u^k\,F(u)\,\diff u,
\end{equation}
and $F$ defined in~\eqref{eq:F_def}.

\subsubsection{Explicit coefficients}

Converting to moments of $\tilde\Delta$ via Fubini
(domain: $Z\leq -u$, $0\leq u\leq -Z$) and using centro-symmetry:
\begin{equation}\label{eq:Ak_Delta}
  A_k
  = \frac{1}{(k+1)!}\int_0^{+\infty} Z^{k+1}\,\tilde\Delta(Z)\,\diff Z.
\end{equation}
The first three coefficients are therefore:
\begin{align}
  A_0 &= \phantom{-}\int_0^{+\infty} Z\,\tilde\Delta(Z)\,\diff Z \;> 0,
  \label{eq:A0}\\[4pt]
  A_1 &= -\frac{1}{2}\int_0^{+\infty} Z^2\,\tilde\Delta(Z)\,\diff Z \;<0,
  \label{eq:A1}\\[4pt]
  A_2 &= \phantom{-}\frac{1}{6}\int_0^{+\infty} Z^3\,\tilde\Delta(Z)\,\diff Z\;>0.
  \label{eq:A2}
\end{align}

\begin{remark}[Sign of $A_1$]
  The coefficient $A_1$ is \emph{negative}: this is a direct consequence
  of the $(-1)^k$ factor in~\eqref{eq:Ak_general} and is crucial for the
  correct sign of the dipolar surface correction.  The signs alternate:
  $A_0>0$, $A_1<0$, $A_2>0$, etc.
\end{remark}

\subsubsection{Distributional limit for higher-order boundary terms}

The same argument applied to each $\chi_n^{\Vmoy}$ defined
in~\eqref{eq:chi_j_bdy} gives:
\begin{equation}\label{eq:distrib_chin}
 \frac{1}{\eta}\,\chi_n^{\Vmoy}\!\left(\frac{z}{\eta}\right)
  \;\xrightarrow[\eta\to 0]{\mathcal{D}'}\;
   \left( \frac{1}{n!}\int_0^{+\infty} Z^{n+1}\,\tilde\Delta(Z)\,\diff Z
 \right) \delta(z).
\end{equation}
The surface coefficient associated with the $n$-th boundary term is
therefore the $(n+1)$-th half-line moment of $\tilde\Delta$, up to
a factorial.

\subsubsection{Physical interpretation of the hierarchy}

The expansion~\eqref{eq:full_expansion} encodes the full hierarchy of
surface effects:
\begin{itemize}
  \item $A_0\,\delta(z)$: the leading \emph{surface susceptibility},
    set by the first moment $\int_0^\infty Z\,\tilde\Delta\,\diff Z$;
  \item $A_1\,\delta'(z)$: a \emph{dipole-layer correction},
    set by the second moment;
  \item $A_2\,\delta''(z)$: a \emph{quadrupole-layer correction},
    set by the third moment.
\end{itemize}
All microscopic complexity of the boundary layer is thereby encoded in
the sequence of half-line moments of $\tilde\Delta$, which constitute the
sole input for the surface electromagnetic problem in this geometry.

\subsubsection{Systematic construction of the kernel}

Conversely, if one wishes to prescribe a given set of surface
coefficients $(A_0, A_1, A_2,\ldots)$, it suffices to choose a kernel
$\tilde\Delta$ with the corresponding prescribed half-line moments.
The most tractable choice is a superposition of exponentials,
\begin{equation}\label{eq:expo_basis}
  \tilde\Delta(Z) = \sum_{k=1}^N a_k\,e^{-|Z|/\ell_k},
\end{equation}
for which all moments are analytically computable: the lengths $\ell_k$
control the nonlocality scales, and the amplitudes $a_k$ fix the moment
conditions.  This representation is the one-dimensional analogue of the
tensor decomposition used in the three-dimensional case
(Section~\ref{sec:3D}).


\section{The three-dimensional tensorial case}
\label{sec:3D}

\subsection{General framework and moment expansion}

\begin{center}
\begin{figure}[h!]
\begin{center}
\includegraphics[scale=0.45]{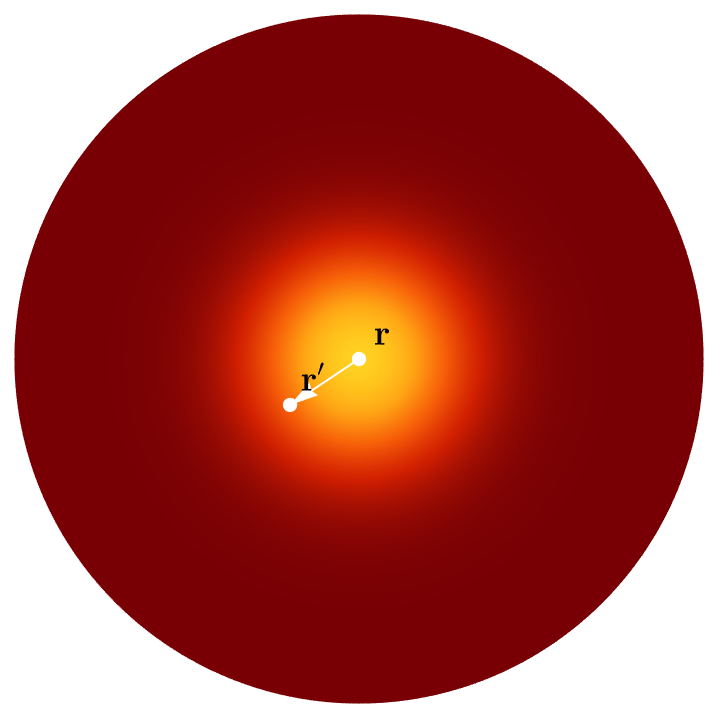}
\includegraphics[scale=0.45]{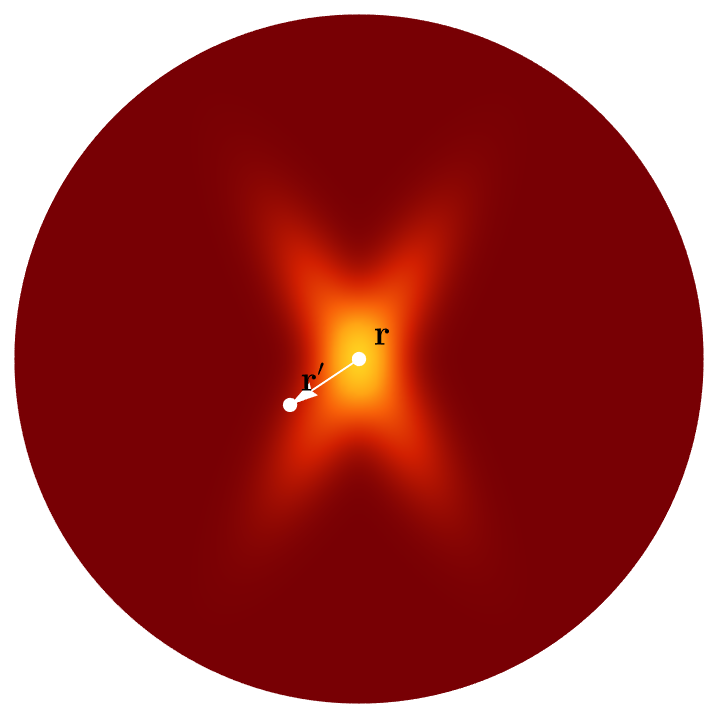}
\end{center}
\caption{Illustration of non-locality for an isotropic material (left) and an anisotropic material (right).\label{fig:iso:aniso}}
\end{figure}
\end{center}

We now extend the analysis to the physically relevant three-dimensional
setting (See Fig. \ref{fig:iso:aniso}).  The most general tensorial nonlocal constitutive relation reads
\begin{equation}\label{eq:constit3D}
  P_i(\br,\omega)
  = \varepsilon_0
  \int_{\R^3}
  \Delta_{ij}^{(1)}(\br,\br',\omega)\,
  E_j(\br',\omega)\,\diff\br',
\end{equation}
where Einstein's summation convention is used throughout and the frequency
dependence $\omega$ is henceforth suppressed.  Setting $\bR := \br'-\br$
and $\diff\br' = \diff\bR$:
\begin{equation}\label{eq:constit3D_R}
  P_i(\br)
  = \varepsilon_0
  \int_{\R^3}
  \Delta_{ij}^{(1)}(\br,\br+\bR)\,
  E_j(\br+\bR)\,\diff\bR.
\end{equation}
Expanding the field about $\br$,
\begin{equation}\label{eq:Taylor3D}
  E_j(\br+\bR)
  = E_j(\br)
  + R_k\,\partial_k E_j(\br)
  + \tfrac{1}{2}R_k R_l\,\partial_k\partial_l E_j(\br)
  + \cdots,
\end{equation}
and inserting into~\eqref{eq:constit3D_R} yields the
\emph{spatial moment expansion}:
\begin{equation}\label{eq:P_moments3D}
  P_i(\br)
  = \varepsilon_0
  \bigl[
    \chi^{(1,0)}_{ij}(\br)\,E_j(\br)
    + \chi^{(1,1)}_{ijk}(\br)\,\partial_k E_j(\br)
    + \chi^{(1,2)}_{ijkl}(\br)\,\partial_k\partial_l E_j(\br)
    + \cdots
  \bigr],
\end{equation}
with the spatial moment tensors
\begin{align}
  \chi^{(1,0)}_{ij}(\br)
  &:= \int_{\R^3}
      \Delta_{ij}^{(1)}(\br,\br+\bR)\,\diff\bR,
  \label{eq:chi0_3D}\\[4pt]
  \chi^{(1,1)}_{ijk}(\br)
  &:= \int_{\R^3}
      R_k\,\Delta_{ij}^{(1)}(\br,\br+\bR)\,\diff\bR,
  \label{eq:chi1_3D}\\[4pt]
  \chi^{(1,2)}_{ijkl}(\br)
  &:= \frac{1}{2}\int_{\R^3}
      R_k R_l\,\Delta_{ij}^{(1)}(\br,\br+\bR)\,\diff\bR.
  \label{eq:chi2_3D}
\end{align}

\subsection{Hypotheses on the kernel}

We impose the following conditions, in direct analogy with the
one-dimensional case.

\begin{enumerate}

\item \textbf{Rapid decay.}
  For every $\br$, the function $\Delta^{(1)}_{ij}(\br,\br+\bR)$
  is rapidly decaying: all tensor moments exist.

\item \textbf{Support in $\Omega\times\Omega$.}
  \begin{equation}
    \Delta^{(1)}_{ij}(\br,\br')
    = \mathds{1}_\Omega(\br)\,\mathds{1}_\Omega(\br')\,
      \Delta^{(1)}_{ij}(\br,\br').
  \end{equation}
  The indicator $\mathds{1}_\Omega(\br)$ ensures zero polarization vector outside
  $\Omega$; $\mathds{1}_\Omega(\br')$ ensures that the exterior region
  does not drive the nonlocal response. 

\item \textbf{Bulk homogeneity.}
  Far from $\dOmega$ the medium is homogeneous:
  \begin{equation}\label{eq:homo3D}
    \Delta^{(1)}_{ij}(\br,\br')
    = \mathds{1}_\Omega(\br)\,\mathds{1}_\Omega(\br')\,
      \tilde\Delta^{(1)}_{ij}(\br-\br').
  \end{equation}

\item \textbf{Bulk isotropy.}
  The bulk kernel is isotropic, i.e., it has the form
  \begin{equation}\label{eq:iso}
    \tilde\Delta^{(1)}_{ij}(\bR)
    = \tilde\Delta_\parallel(R)\,\delta_{ij}
    + \bigl[\tilde\Delta_\perp(R)-\tilde\Delta_\parallel(R)\bigr]
      \frac{R_i R_j}{R^2},
    \quad R = |\bR|,
  \end{equation}
  where $\tilde\Delta_\parallel$ and $\tilde\Delta_\perp$ are two scalar
  radial functions.  This is the most general form of a rank-2
  tensor-valued function of $\bR$ that is covariant under all
  rotations; see~\cite{Agranovich1984,LandauLifshitz1984} and
  Appendix~\ref{app:isotropy} for a self-contained proof.

\item \textbf{Centro-symmetry.}
  $\tilde\Delta^{(1)}_{ij}(-\bR) = \tilde\Delta^{(1)}_{ij}(\bR)$,
  which is automatically satisfied by~\eqref{eq:iso}.

\end{enumerate}

\begin{remark}
  The assumption $\mathds{1}_\Omega(\br')$ in Hypothesis~2 implies that
  the nonlocal interaction is cut off sharply at the interface.  While
  quantum mechanics tells us that the actual transition is smooth over an
  atomic scale, this hypothesis does not simplify the calculations
  significantly and can be relaxed at the cost of additional terms in the
  moment expansion.  We retain it here for clarity.
\end{remark}

\begin{remark}
The last two assumptions are not strictly necessary and could be lifted
if one were to study materials that are either anisotropic or
non-centrosymmetric.  The non-centrosymmetric case, in which the
rank-one moment $\chi^{(1,1)}_{ijk}$ no longer vanishes, is the
subject of the companion paper~\cite{ArticleChiral}, where it is shown
to lead to optical activity and the Drude--Born--Fedorov constitutive
relation.
\end{remark}

\subsection{Bulk and boundary decomposition}
\begin{figure}[h!]
\begin{center}
\includegraphics[scale=0.4]{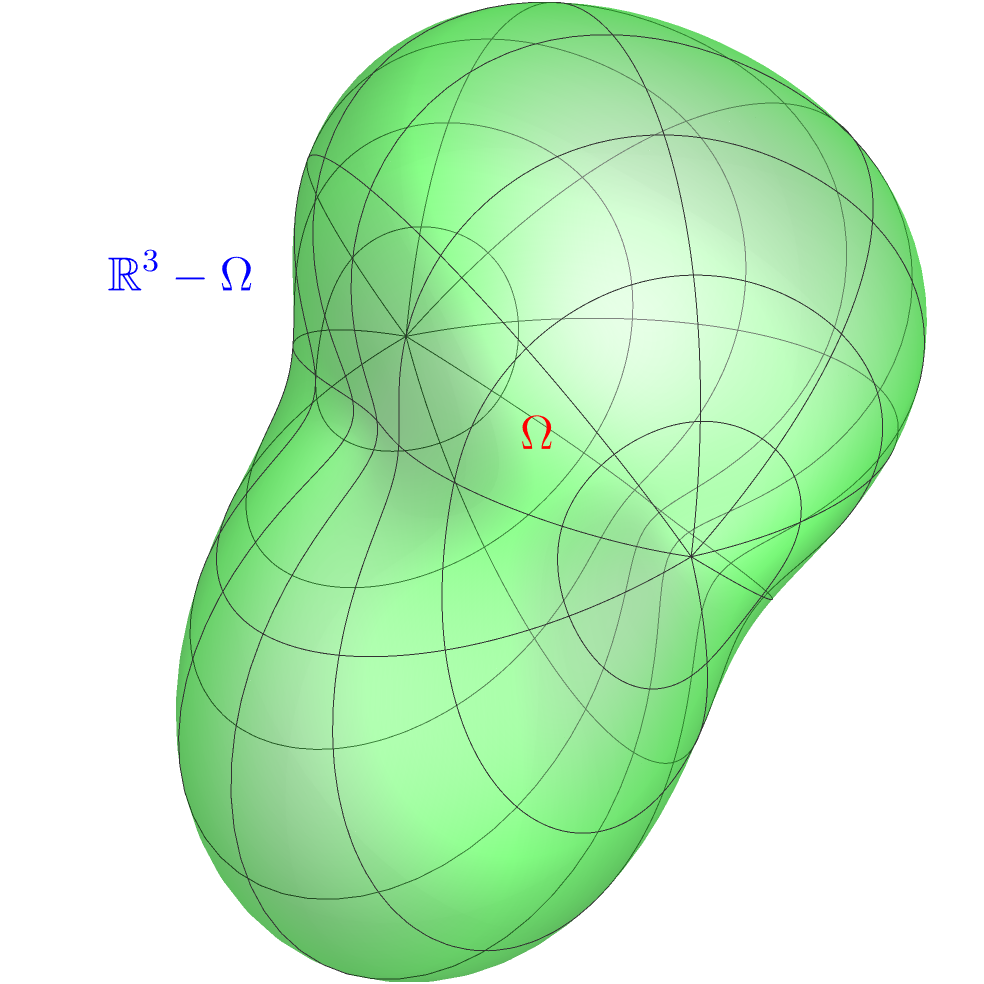}
\caption{The set $\Omega$ is supposed to be filled with homogeneous but non-local material. $\Omega^c = \mathbb{R}^3 -\Omega$, on the other hand, is assumed to be filled with a local medium. \label{fig:potato}}
\end{center}
\end{figure}

In the sequel, the set $\Omega$ is supposed to be a simply connected and sufficiently smooth domain (See Fig. \ref{fig:potato}) and $\Omega^c$ denotes the complement of $\Omega$ in $\mathbb{R}^3$.

. 
For $\br\in\Omega$, we decompose the integration domain as
$\R^3 = (\Omega-\br)\cup(\Omega^c-\br)$, where
$\Omega-\br := \{\bR;\,\br+\bR\in\Omega\}$, and write
\begin{equation}
  \int_{\R^3}\tilde\Delta^{(1)}_{ij}(\bR)\,\diff\bR
  = \int_{\Omega-\br}\tilde\Delta^{(1)}_{ij}(\bR)\,\diff\bR
  + \int_{\Omega^c-\br}\tilde\Delta^{(1)}_{ij}(\bR)\,\diff\bR.
\end{equation}
Subtracting and adding the full-space integral, the polarization
\eqref{eq:constit3D_R} splits into a bulk part and an interfacial
correction:
\begin{equation}\label{eq:decomp_P3D}
  P_i(\br) = P_i^{\Omega}(\br) + P^{\Vmoy}_i(\br),
\end{equation}
where
\begin{equation}\label{eq:Pbulk3D}
  P_i^{\Omega}(\br)
  := \varepsilon_0\,\mathds{1}_\Omega(\br)
     \int_{\R^3}
     \tilde\Delta^{(1)}_{ij}(\bR)\,E_j(\br+\bR)\,\diff\bR
\end{equation}
is the \emph{bulk polarization} (that of an infinite homogeneous medium),
and
\begin{equation}\label{eq:deltaP3D}
  P^{\Vmoy}_i(\br)
  := -\varepsilon_0\,\mathds{1}_\Omega(\br)
     \int_{\Omega^c}
     \tilde\Delta^{(1)}_{ij}(\br-\br')\,E_j^+(\br')\,\diff\br'
\end{equation}
is the \emph{interfacial contribution}, concentrated in a layer of
thickness $\ell$ near $\dOmega$.  Here $E_j^+$ denotes the field value
on the exterior side $\Omega^c$.

Accordingly, each spatial moment tensor decomposes as
\begin{equation}\label{eq:decomp_chi3D}
  \chi^{(1,n)}_{ij\,k_1\cdots k_n}(\br)
  = \chi^{\Omega,(n)}_{ij\,k_1\cdots k_n}
  - \chi^{\Vmoy,(n)}_{ij\,k_1\cdots k_n}(\br),
\end{equation}
with the constant \emph{bulk moment}
\begin{equation}\label{eq:chi_bulk3D}
  \chi^{\Omega,(n)}_{ij\,k_1\cdots k_n}
  := \frac{1}{n!}
     \int_{\R^3}
     R_{k_1}\cdots R_{k_n}\,
     \tilde\Delta^{(1)}_{ij}(\bR)\,\diff\bR,
\end{equation}
and the \emph{boundary term}
\begin{equation}\label{eq:chi_bdy3D}
  \chi^{\Vmoy,(n)}_{ij\,k_1\cdots k_n}(\br)
  := \frac{\mathds{1}_\Omega(\br)}{n!}
     \int_{\Omega^c-\br}
     R_{k_1}\cdots R_{k_n}\,
     \tilde\Delta^{(1)}_{ij}(\bR)\,\diff\bR.
\end{equation}

\paragraph{Vanishing of odd bulk moments.}
By bulk isotropy~\eqref{eq:iso} and centro-symmetry,
$\bR\mapsto R_{k_1}\cdots R_{k_n}\tilde\Delta^{(1)}_{ij}(\bR)$ is odd
for odd $n$, so $\chi^{\Omega,(n)} = 0$ for all odd $n$.
In particular $\chi^{\Omega,(1)}_{ijk}=0$, and the first nonlocal bulk
correction appears at order $n=2$.

\begin{remark}[$E^+$ vs. $E^-$]
  The tangential component of $\bE$ does not suffer a jump when crossing $\dOmega$
  ($\bE^+_\parallel = \bE^-_\parallel$), while the normal component is
  generally discontinuous.  Since the integral \eqref{eq:deltaP3D} is
  taken over $\Omega^c$, it is $E^+_j$ that appears naturally; the
  interior limit $E^-_j$ does not enter this term.  This asymmetry
  is structural, not imposed by hand.
\end{remark}

\subsection{Interface geometry}

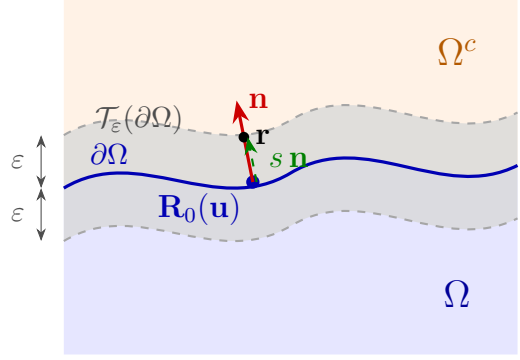
\begin{figure}[ht]
  \centering
  \begin{tikzpicture}[>=Stealth, thick]
    \def\eps{0.7}
    \def\surface{(0,0) .. controls (1,0.6) and (2,-0.4) .. (3,0.2)
                       .. controls (4,0.8) and (5,-0.2) .. (6,0.3)}
    \fill[blue!10] \surface -- (6,-2.2) -- (0,-2.2) -- cycle;
    \fill[orange!10] \surface -- (6,2.5) -- (0,2.5) -- cycle;
    \fill[gray!30, opacity=0.8]
      (0,\eps) .. controls (1,0.6+\eps) and (2,-0.4+\eps) .. (3,0.2+\eps)
               .. controls (4,0.8+\eps) and (5,-0.2+\eps) .. (6,0.3+\eps)
      -- (6,0.3-\eps) .. controls (5,-0.2-\eps) and (4,0.8-\eps) .. (3,0.2-\eps)
                      .. controls (2,-0.4-\eps) and (1,0.6-\eps) .. (0,-\eps)
      -- cycle;
    \draw[dashed, gray!70]
      (0,\eps) .. controls (1,0.6+\eps) and (2,-0.4+\eps) .. (3,0.2+\eps)
               .. controls (4,0.8+\eps) and (5,-0.2+\eps) .. (6,0.3+\eps);
    \draw[dashed, gray!70]
      (0,-\eps) .. controls (1,0.6-\eps) and (2,-0.4-\eps) .. (3,0.2-\eps)
                .. controls (4,0.8-\eps) and (5,-0.2-\eps) .. (6,0.3-\eps);
    \draw[blue!70!black, very thick] \surface;
    \coordinate (R0) at (2.5, 0.08);
    \fill[blue!70!black] (R0) circle (2.5pt);
    \node[below left, blue!70!black] at (R0) {$\mathbf{R}_0(\mathbf{u})$};
    \def\nx{-0.196}\def\ny{0.981}
    \def\nlen{1.1}
    \coordinate (nend) at ($(R0) + \nlen*(\nx,\ny)$);
    \draw[->, red!80!black, very thick] (R0) -- (nend)
      node[right, red!80!black] {$\mathbf{n}$};
    \def\s{0.55}
    \coordinate (rpt) at ($(R0) + \s*\nlen*(\nx,\ny)$);
    \fill[black] (rpt) circle (2pt);
    \node[right] at (rpt) {$\mathbf{r}$};
    \draw[->, green!50!black, thick, dashed]
      ($(R0)+(0.05,0)$) -- ($(rpt)+(0.05,0)$)
      node[midway, right=2pt, green!50!black] {$s\,\mathbf{n}$};
    \draw[<->, gray!60!black, thin]
      (-0.3,0) -- (-0.3,\eps)
      node[midway, left=2pt, gray!60!black, font=\small] {$\varepsilon$};
    \draw[<->, gray!60!black, thin]
      (-0.3,0) -- (-0.3,-\eps)
      node[midway, left=2pt, gray!60!black, font=\small] {$\varepsilon$};
    \node[blue!60!black, font=\large] at (5.2,-1.4) {$\Omega$};
    \node[orange!70!black, font=\large] at (5.2, 1.8) {$\Omega^c$};
    \node[gray!60!black, font=\small] at (1.0, 0.9)
      {$\mathcal{T}_\varepsilon(\partial\Omega)$};
    \node[blue!70!black, font=\small] at (0.6, 0.45) {$\partial\Omega$};
  \end{tikzpicture}
  \caption{Tubular neighbourhood $\mathcal{T}_\varepsilon(\partial\Omega)$
    (grey band) of width $\varepsilon$ around the interface $\partial\Omega$
    (blue curve).  Every point $\mathbf{r}$ in the tube admits a unique
    decomposition $\mathbf{r} = \mathbf{R}_0(\mathbf{u}) + s\,\mathbf{n}$,
    where $s\in(-\varepsilon,\varepsilon)$ is the signed distance to
    $\partial\Omega$ (positive in $\Omega^c$, negative in $\Omega$).
    The thin-layer limit $\varepsilon\to 0$ concentrates all interfacial
    contributions onto $\partial\Omega$.}
  \label{fig:tubular}
\end{figure}

Let $\dOmega$ be a smooth surface parametrized by $\bR_0(\bu)$,
$\bu=(u_1,u_2)$.  The \emph{tubular neighbourhood} of $\dOmega$ of
width $\varepsilon$ is the open set
\begin{equation}
  \mathcal{T}_\varepsilon(\dOmega)
  := \bigl\{\br\in\R^3 : \mathrm{dist}(\br,\dOmega) < \varepsilon\bigr\}.
\end{equation}
Provided $\varepsilon$ is smaller than the smallest radius of curvature
of $\dOmega$, every point $\br\in\mathcal{T}_\varepsilon(\dOmega)$
admits a \emph{unique} decomposition
\begin{equation}
  \br = \bR_0(\bu) + s\,\bn(\bu),
\end{equation}
where $\bn$ is the outward unit normal and $s\in(-\varepsilon,\varepsilon)$
is the \emph{signed distance} to $\dOmega$, positive in $\Omega^c$ and
negative in $\Omega$ (see Figure~\ref{fig:tubular} and~\cite{Gray2004}).  The
volume element in normal coordinates is
\begin{equation}\label{eq:jacobian}
  \diff\br
  = \bigl(1 + s H + s^2 K + O(s^3)\bigr)\,\diff S\,\diff s,
\end{equation}
where $\diff S$ is the surface area element, $H = \kappa_1+\kappa_2$
the mean curvature, and $K = \kappa_1\kappa_2$ the Gaussian curvature (See Annex \ref{app:jacobian}, p. \pageref{app:jacobian}).

We decompose $\bR = \br'-\br$ into normal and tangential parts:
\begin{equation}
  \bR = \bR_\parallel + \Rp\,\bn,
  \qquad
  \Rp := \bR\cdot\bn,
  \qquad
  \bR_\parallel := \bR - \Rp\,\bn.
\end{equation}

\subsection{Distributional thin-layer limit and surface susceptibilities}
\label{sec:distrib3D}

The boundary terms~\eqref{eq:chi_bdy3D} are concentrated in a layer of
thickness $\ell$ around $\dOmega$.  We apply the distributional
thin-layer procedure of Section~\ref{sec:1D_distrib}, now in three
dimensions with the Jacobian~\eqref{eq:jacobian}.

\subsubsection{Leading-order moment ($M_0$)}

At leading order in $\ell/\lambda$, the thin-layer limit of
$\chi^{\Vmoy,(n)}_{ij\,k_1\cdots k_n}$ yields a distribution supported
on $\dOmega$:
\begin{equation}\label{eq:M0_tensor}
  \chi^{\Vmoy,(n)}_{ij\,k_1\cdots k_n}(\br)
  \;\xrightarrow{\ell\to 0}\;
  \mathcal{M}^{(n)}_{ij\,k_1\cdots k_n}\,\delta_{\dOmega},
\end{equation}
with the \emph{surface moment}
\begin{equation}\label{eq:M0_def}
  \mathcal{M}^{(n)}_{ij\,k_1\cdots k_n}(\bR_0)
  := \frac{1}{n!}
     \int_0^{+\infty}\int_{\R^2}
     R_{k_1}\cdots R_{k_n}\,
     \tilde\Delta^{(1)}_{ij}(\bR)\,
     \diff^2\bR_\parallel\,\diff\Rp,
\end{equation}
where the half-space $\Rp > 0$ corresponds to the exterior $\Omega^c$
near $\dOmega$.

\subsubsection{Vanishing of the first-order correction ($M_1 = 0$)}

The first-order correction in $\ell$ arises from the linear term $sH$
in the Jacobian expansion~\eqref{eq:jacobian}.  Concretely, it
contributes an integral of the form
\begin{equation}\label{eq:M1_integral}
  \mathcal{M}^{(1)}_{ij}
  \;\propto\;
  H \int_{-\infty}^{+\infty} s\,\diff s
  \int_{\R^2}
  \tilde\Delta^{(1)}_{ij}(\bR_\parallel, s)\,\diff^2\bR_\parallel,
\end{equation}
where the integration runs over the thin layer $|s|\lesssim\ell$ and
over the tangential displacement $\bR_\parallel$.  For a bulk-isotropic
kernel~\eqref{eq:iso}, the integrand is \emph{even} in $s$ (since
$\tilde\Delta_{ij}$ depends only on $R = |\bR|= \sqrt{R_\parallel^2+s^2}$),
while the prefactor $s$ is \emph{odd}.  The full integrand is therefore
odd under $s\to -s$, and integrates to zero on the symmetric domain
$s\in(-\ell,+\ell)$:
\begin{equation}\label{eq:M1_zero}
  \mathcal{M}^{(1)}_{ij\,k_1\cdots k_n} = 0.
\end{equation}
This is a fundamental result: for an isotropic bulk kernel, the curvature
corrections $\partial_n\delta_{\dOmega}$ and $H\,\delta_{\dOmega}$
at order $\ell$ both vanish simultaneously.  Curvature effects first
appear at order $\ell^2$.

\subsubsection{Full distributional expansion}

Combining the above, the interfacial polarization~\eqref{eq:deltaP3D}
admits the distributional expansion:
\begin{equation}\label{eq:deltaP_distrib3D}
  P^{\partial \Omega}_i(\br)
  = -\varepsilon_0
  \Bigl[
    \mathcal{S}^{(0)}_{ij}\,\delta_{\dOmega}
    + \mathcal{S}^{(1)}_{ijk}\,\partial_k\delta_{\dOmega}
    + \ell^2\,\mathcal{S}^{(2)}_{ij}\,H\,\delta_{\dOmega}
    + \ell^2\,\mathcal{S}^{(3)}_{ij}\,K\,\delta_{\dOmega}
    + \cdots
  \Bigr]
  E_j^+
  + O(\ell^3/\lambda^3),
\end{equation}
where $\mathcal{S}^{(1)}_{ijk} = 0$ by isotropy, and the remaining
tensors $\mathcal{S}^{(p)}$ are given by moments of
$\tilde\Delta^{(1)}_{ij}$ over the exterior half-space.
\begin{remark}
It is isotropy and centrosymmetry \emph{together} that lead to the vanishing of the
term $\mathcal{S}^{(1)}_{ijk}$, for if the centrosymmetry constraint is
relaxed, isotropic tensors of order~3 do exist. They are then proportional to
the Levi-Civita pseudo-tensor $\epsilon_{ijk}$, which is completely
antisymmetric, and the tensor takes the form
$\mathcal{S}^{(1)}_{ijk} = S_0\,\epsilon_{ijk}$.
The scalar $S_0$ is a \emph{pseudo-scalar}: it changes sign under
spatial inversion, and its non-vanishing signals the chirality of the
medium.  Since $\epsilon_{ijk}\,\partial_j E_k \propto (\nabla\times\mathbf{E})_i
\propto B_i$, this term introduces a coupling between the electric
polarization and the magnetic field, which is precisely the
Drude--Born--Fedorov constitutive relation characteristic of optically
active media.  This non-centrosymmetric regime (isotropic but chiral !)
is the subject of the companion paper~\cite{ArticleChiral}.
\end{remark}

\subsubsection{Surface susceptibility}

By bulk isotropy~\eqref{eq:iso} and centrosymmetry, the half-space
integral $\mathcal{S}^{(0)}_{ij}$ is proportional to $\delta_{ij}$
(proved in Appendix~\ref{app:chi_derivation}):
\begin{equation}\label{eq:S0_decomp}
  \mathcal{S}^{(0)}_{ij} = \chis\,\delta_{ij},
\end{equation}
with the single \emph{surface susceptibility}
\begin{equation}\label{eq:chis3D}
  \chis
  := \frac{2\mu_0^\parallel + \mu_0^\perp}{6},
\end{equation}
and the scalar half-line moments
\begin{equation}\label{eq:mu_n}
  \mu_n^\alpha
  := 4\pi\int_0^{+\infty} r^{n+2}\,\tilde\Delta_\alpha(r)\,\diff r,
  \qquad \alpha\in\{\parallel,\perp\},\quad n\in\mathbb{N}.
\end{equation}
The scalar $\chis$ is the \emph{surface susceptibility}: it encodes the
isotropic response of the interfacial layer and constitutes a
constructive generalization of the Feibelman $d$-parameters
(see section~\ref{sec:CL}).

\begin{remark}
  For a centrosymmetric isotropic kernel the tangential and normal
  responses are \emph{equal}: $\chispar = \chisper = \chis$.  The
  notation $\chispar \neq \chisper$ only becomes meaningful when either
  isotropy or centrosymmetry is broken (cf.\ the chiral
  companion paper~\cite{ArticleChiral}).
\end{remark}

\subsubsection{Curvature corrections}

At order $\ell^2$, the coefficients of $H\,\delta_{\dOmega}$ and
$K\,\delta_{\dOmega}$ in~\eqref{eq:deltaP_distrib3D} are given by the
second moments of the kernel:
\begin{equation}\label{eq:curv_coeff}
  \nu^\parallel
  := \frac{2\pi}{3}\,\mu_2^\parallel,
  \qquad
  \nu^\perp
  := \frac{2\pi}{3}\,\mu_2^\perp.
\end{equation}
These corrections are responsible for the size-dependent optical
properties of curved nanostructures (Mie corrections) and will be
evaluated explicitly for each geometry below.

\subsection{Hierarchy of surface contributions}

The following Table~\ref{tab:hierarchy} summarises the distributional expansion.

\begin{table}[ht]
\centering
\renewcommand{\arraystretch}{1.6}
\begin{tabular}{|c|l|l|l|}
\toprule
Order & Distributional term & Coefficient & Physical meaning \\
\midrule
$\ell^0$ & $\delta_{\dOmega}$
  & $\chis$
  & Surface susceptibility \\
$\ell^1$ & $\partial_n\delta_{\dOmega}$,\; $H\,\delta_{\dOmega}$
  & $0$
  & Vanishes by isotropy \\
$\ell^2$ & $H\,\delta_{\dOmega}$
  & $\nu^\parallel,\;\nu^\perp$
  & Mean-curvature correction \\
$\ell^2$ & $K\,\delta_{\dOmega}$
  & $\sim\mu_2^\parallel,\;\mu_2^\perp$
  & Gaussian-curvature (topological) \\
\bottomrule
\end{tabular}
\caption{Hierarchy of distributional surface contributions in the thin-layer
  expansion of the isotropic centrosymmetric nonlocal response. Each row
  gives the order in the kernel decay length~$\ell$, the associated
  distributional term supported on~$\dOmega$, the corresponding scalar
  coefficient and its physical interpretation.}
\label{tab:hierarchy}
\end{table}

\begin{remark}[Universality]
  This hierarchy is \emph{universal}: it depends on the microscopic
  kernel $\tilde\Delta^{(1)}_{ij}$ only through its moments, and on the
  surface $\dOmega$ only through its geometric invariants $H$ and $K$.
  The same structure extends to nonlinear response, as shown in the
  companion paper~\cite{ArticleNL}.
\end{remark}

\subsection{Special geometries}
\label{sec:geometries}

We now instantiate the general result for four canonical geometries.

\subsubsection{Planar interface ($H=K=0$)}

For $\dOmega = \{z=0\}$ with $\Omega = \{z>0\}$ and outward normal
$\bn = -\hat{\mathbf{z}}$, both curvatures vanish and the Jacobian
in~\eqref{eq:jacobian} equals $1$ identically.  The
explicit angular integration (see Appendix~\ref{app:chi_derivation})
confirms~\eqref{eq:chis3D}.  The interfacial polarization reduces to:
\begin{equation}\label{eq:deltaP_plan}
  P^{\partial \Omega}_i(\br)
  = -\varepsilon_0\,\chis\,E_i^+\,\delta(z)
  + O(\ell^2/\lambda^2).
\end{equation}
There is neither a dipole-layer term nor a curvature correction:
the planar interface is fully characterised by the single scalar~$\chis$.

\subsubsection{Spherical interface}

For the sphere of radius $R_0$:
$\kappa_1=\kappa_2 = 1/R_0$, $H = 2/R_0$, $K = 1/R_0^2$,
$\bn = \hat{\mathbf{r}}$.  The Jacobian is exact:
$(1+s/R_0)^2 = 1 + 2s/R_0 + s^2/R_0^2$.

At leading order the surface susceptibility $\chis$ is identical to
that of the planar case.  The curvature correction
at order $\ell^2/R_0$ gives:
\begin{equation}\label{eq:deltaP_sphere}
  P^{\partial \Omega}_i(\br)
  = -\varepsilon_0
  \left[
    \mathcal{S}^{(0)}_{ij}
    + \frac{2}{R_0}
    \bigl(\nu^\parallel P^\parallel_{ij}
          + \nu^\perp P^\perp_{ij}\bigr)
  \right]
  \delta(r-R_0)\,E_j^+
  + O\!\left(\frac{\ell^2}{R_0^2}\right).
\end{equation}
The term proportional to $\ell^2/R_0$ is the \emph{nonlocal Mie
correction}: it shifts the resonance frequencies of plasmonic
nanospheres and becomes observable for $R_0\lesssim 10\ell$ (See paragraph \ref{paraph:Curv:on-a-sphere}, for explanations).  In the
limit $R_0\to\infty$ one recovers~\eqref{eq:deltaP_plan} as required.

\subsubsection{Cylindrical interface}

For the cylinder of radius $R_0$ with axis $\hat{\mathbf{z}}$:
$\kappa_1 = 1/R_0$ (azimuthal direction),
$\kappa_2 = 0$ (axial direction),
$H = 1/R_0$, $K = 0$.  The Jacobian is exact: $1 + s/R_0$.

\begin{equation}\label{eq:deltaP_cylinder}
  P^{\partial \Omega}_i(\br)
  = -\varepsilon_0
  \left[
    \mathcal{S}^{(0)}_{ij}
    + \frac{1}{R_0}
    \bigl(\nu^\parallel P^\parallel_{ij}
          + \nu^\perp P^\perp_{ij}\bigr)
  \right]
  \delta(\rho-R_0)\,E_j^+
  + O\!\left(\frac{\ell^2}{R_0^2}\right),
\end{equation}
where $\rho = \sqrt{x^2+y^2}$.  The mean-curvature correction is half
that of the sphere of equal radius, reflecting the fact that the cylinder
curves in only one tangential direction.  Since $K=0$, no topological
correction appears.

\begin{remark}[Induced surface anisotropy]
  For the cylinder, the two tangential directions, azimuthal
  $\hat{\boldsymbol\varphi}$ (curved) and axial $\hat{\mathbf{z}}$
  (flat), are inequivalent even though the bulk kernel is isotropic.
  This induces an effective surface anisotropy that can be resolved at
  order $\ell^2/R_0^2$ by decomposing $P^\parallel_{ij}$ onto the
  azimuthal and axial components.
\end{remark}

\subsubsection{Ellipsoidal interface}

For the ellipsoid with semi-axes $a$, $b$, $c$ parametrized by
$(\theta,\varphi)$, both $H$ and $K$ vary over the surface.  The general
distributional expansion~\eqref{eq:deltaP_distrib3D} holds pointwise,
with $\mathcal{S}^{(0)}_{ij}(\bR_0)$ depending on position through the
local normal $\bn(\bR_0)$:
\begin{equation}\label{eq:deltaP_ellipsoid}
 P^{\partial \Omega}_i(\br)
  = -\varepsilon_0
  \Bigl[
    \mathcal{S}^{(0)}_{ij}(\bR_0)\,\delta_{\dOmega}
    + \ell^2
    \bigl(\nu^\parallel H(\bR_0)\,P^\parallel_{ij}(\bR_0)
         + \nu^\perp H(\bR_0)\,P^\perp_{ij}(\bR_0)\bigr)
    \delta_{\dOmega}
  \Bigr]E_j^+
  + O(\ell^3).
\end{equation}
The surface susceptibilities are locally anisotropic: they vary from
point to point on the ellipsoid through the direction of $\bn(\bR_0)$.

The following limiting cases are recovered immediately:
\begin{itemize}
  \item $a=b=c=R_0$ (sphere): $H=2/R_0$, $K=1/R_0^2$,
    $\bn=\hat{\mathbf{r}}$ — case~B above.
  \item $c\to\infty$, $a=b$ fixed (cylinder): $H\to 1/a$, $K\to 0$
    — case~C above.
  \item $c\to 0$, $a=b$ fixed (flat disk): $H\to\infty$ at the rim,
    signalling the geometric singularity of the edge.
\end{itemize}

\begin{table}[ht]
\centering
\small
\renewcommand{\arraystretch}{1.5}
\begin{tabular}{|l|c|c|c|p{4.2cm}|}
\toprule
Case & $H$ & $K$ & Exact Jacobian & Feature \\
\midrule
A: Plane
  & $0$ & $0$ & $1$
  & Reference case; no curvature corrections \\
B: Sphere $R_0$
  & $2/R_0$ & $1/R_0^2$ & $(1+s/R_0)^2$
  & Isotropic Mie correction \\
C: Cylinder $R_0$
  & $1/R_0$ & $0$ & $1+s/R_0$
  & Anisotropic; correction halved vs.\ sphere \\
D: Ellipsoid
  & variable & variable & complex
  & Locally anisotropic surface susceptibility \\
\bottomrule
\end{tabular}
\caption{Geometric invariants (mean curvature~$H$, Gaussian curvature~$K$)
  and exact thin-shell Jacobian for the four reference surfaces treated
  analytically. The rightmost column highlights the distinctive feature of
  each geometry with respect to the surface susceptibility.}
\label{tab:geometries}
\end{table}

\subsection{Explicit kernel/surface combinations}
\label{sec:explicit}

We treat four combinations for which all calculations can be carried
through analytically.  We recall the decomposition~\eqref{eq:iso} and
the scalar moments~\eqref{eq:mu_n}.

\subsubsection{Gaussian scalar kernel on a planar interface}

\paragraph{Kernel.}
$\tilde\Delta^{(1)}_{ij}(\bR) = \Delta_0\,e^{-R^2/\ell^2}\,\delta_{ij}$,
i.e., $\tilde\Delta_\parallel = \tilde\Delta_\perp =
\Delta_0\,e^{-R^2/\ell^2}$.

\paragraph{Scalar moments.}
Setting $u = R/\ell$:
\begin{equation}
  \mu_n = 4\pi\Delta_0\ell^3
  \int_0^{+\infty} u^{n+2}\,e^{-u^2}\,\diff u
  = 4\pi\Delta_0\ell^3 \times \tfrac{1}{2}\,\Gamma\!\left(\tfrac{n+3}{2}\right).
\end{equation}
In particular:
\begin{equation}
  \mu_0 = \Delta_0\,\pi^{3/2}\ell^3,
  \qquad
  \mu_2 = \tfrac{3}{2}\,\mu_0\,\ell^2.
\end{equation}

\paragraph{Surface susceptibility (plane, $H=K=0$).}
Since $\tilde\Delta_\parallel = \tilde\Delta_\perp$,
equation~\eqref{eq:chis3D} gives:
\begin{equation}\label{eq:chi_gauss}
  \chis = \frac{3\mu_0}{6} = \frac{\mu_0}{2}
  = \frac{\Delta_0\,\pi^{3/2}\ell^3}{2}.
\end{equation}
The factor $1/2$ arises because the half-space integral covers only
$\Rp > 0$.

\paragraph{Interfacial polarization.}
\begin{equation}
  P^{\partial \Omega}_i = -\frac{\varepsilon_0\Delta_0\pi^{3/2}\ell^3}{2}\,
  \delta_{ij}\,E_j^+\,\delta(z)
  + O(\ell^2/\lambda^2).
\end{equation}

\subsubsection{Yukawa scalar kernel on a spherical interface}

\paragraph{Kernel.}
$\tilde\Delta^{(1)}_{ij}(\bR) = \Delta_0\ell\,e^{-R/\ell}/R\, \delta_{ij}$,
i.e., $\tilde\Delta_\parallel = \tilde\Delta_\perp = \Delta_0\ell\,e^{-R/\ell}/R$.

\paragraph{Scalar moments.}
\begin{equation}
  \mu_0 = 4\pi\Delta_0\ell^3,
  \qquad
  \mu_2 = 6\ell^2\,\mu_0 = 24\pi\Delta_0\ell^5.
\end{equation}

\paragraph{Surface susceptibilities.}
$\chispar = \chisper = \mu_0/2 = 2\pi\Delta_0\ell^3$.

\paragraph{Curvature correction on the sphere.}\label{paraph:Curv:on-a-sphere}
From~\eqref{eq:deltaP_sphere} with $H = 2/R_0$ and
$\nu^{\parallel,\perp} = \frac{2\pi}{3}\mu_2$:
\begin{equation}\label{eq:mie_yukawa}
  \frac{|\delta P^{\text{curv}}|}{|\delta P^{(0)}|}
  = \frac{(4\pi/3)\times 24\pi\Delta_0\ell^5/R_0}
         {2\pi\Delta_0\ell^3}
  = \frac{16\pi\ell^2}{R_0}.
\end{equation}
For $R_0 = 10\,\mathrm{nm}$ and $\ell = 0.3\,\mathrm{nm}$, this gives
$\approx 1.5\times10^{-2}$: the Mie correction is of order $1\%$,
observable in nanophotonic experiments.

\subsubsection{Tensorial Lorentz kernel on a planar interface}

\paragraph{Kernel.}
\begin{equation}\label{eq:lorentz_kernel}
  \tilde\Delta^{(1)}_{ij}(\bR)
  = \frac{\Delta_0}{(1+R^2/\ell^2)^2}
  \left[\delta_{ij}
        + \frac{\beta\ell^2}{R^2+\ell^2}\,\frac{R_i R_j}{R^2}
  \right],
\end{equation}
with $\tilde\Delta_\parallel = \Delta_0/(1+R^2/\ell^2)^2$ and
$\tilde\Delta_\perp = \tilde\Delta_\parallel(1+\beta\ell^2/(R^2+\ell^2))$.
The parameter $\beta$ controls the normal/tangential anisotropy of the bulk
kernel; $\beta=0$ recovers the isotropic case.

\begin{remark}[Illustrative model]
  Equation~\eqref{eq:lorentz_kernel} is not derived from a specific microscopic
  model but is chosen for its analytical tractability: the moments
  $\mu_0^{\parallel}$, $\mu_0^{\perp}$, and the surface susceptibility
  $\chi^s$ all follow in closed form.  The scalar envelope
  $(1+R^2/\ell^2)^{-2}$ is the three-dimensional Cauchy--Lorentz profile,
  which decays as $R^{-4}$ and is square-integrable.  The tensorial factor
  $\beta\ell^2/(R^2+\ell^2)$ is the simplest choice that (i)~respects
  SO(3) invariance, (ii)~introduces a finite longitudinal/transverse
  anisotropy at $R=0$, and (iii)~vanishes at large $R$, so that the kernel
  is asymptotically isotropic.  The parameter $\beta$ is not fixed by
  symmetry alone; its value is determined by the underlying microscopic
  model.
\end{remark}

\paragraph{Scalar moments (setting $u=R/\ell$ and using
$\int_0^\infty u^2/(1+u^2)^2\,\diff u = \pi/4$,
$\int_0^\infty u^2/(1+u^2)^3\,\diff u = \pi/8$):}
\begin{equation}
  \mu_0^\parallel = \pi^2\Delta_0\ell^3,
  \qquad
  \mu_0^\perp = \pi^2\Delta_0\ell^3\!\left(1+\frac{\beta}{2}\right).
\end{equation}

\paragraph{Surface susceptibility (plane).}
Inserting into~\eqref{eq:chis3D}:
\begin{equation}\label{eq:chis_lorentz}
  \chis
  = \frac{2\mu_0^\parallel+\mu_0^\perp}{6}
  = \frac{\pi^2\Delta_0\ell^3(6+\beta)}{12}.
\end{equation}

\begin{remark}[No surface anisotropy for SO(3)-invariant kernels]
  Although $\tilde\Delta_\parallel \neq \tilde\Delta_\perp$ when $\beta\neq 0$,
  the half-space integral still gives $\mathcal{S}^{(0)}_{ij}\propto\delta_{ij}$
  (see Appendix~\ref{app:chi_derivation}).  The reason is that the
  $\varphi$ integration yields equal angular factors for the tangential
  ($\pi\cdot\frac{4}{3}$) and normal ($2\pi\cdot\frac{2}{3}$) components,
  cancelling the apparent asymmetry between $\tilde\Delta_\parallel$ and
  $\tilde\Delta_\perp$.  Surface anisotropy ($\chispar\neq\chisper$) can
  only arise from kernels that break SO(3) invariance, e.g., through
  interface-induced anisotropy not captured by the present bulk model.
\end{remark}

\subsubsection{Local limit: recovery of Fresnel conditions}

\paragraph{Kernel.}
$\tilde\Delta^{(1)}_{ij}(\bR) = \chi_0\,\delta^{(3)}(\bR)\,\delta_{ij}$. We are then in the case of a local material.
All moments of order $n\geq 1$ vanish: $\mu_n = 0$ for $n\geq 1$.

\paragraph{Result (arbitrary surface geometry).}
For any surface $\dOmega$, the curvature corrections are all zero
($\nu^\parallel = \nu^\perp = 0$), and the interfacial polarization
reduces to:
\begin{equation}\label{eq:deltaP_local}
  P^{\partial \Omega}_i(\br)
  = -\frac{\varepsilon_0\chi_0}{2}\,E_i^+\,\delta_{\dOmega},
\end{equation}
regardless of $H$ and $K$.  This is the distributional form of the
\emph{Fresnel conditions}: the interface is a simple susceptibility
discontinuity, with no internal structure.  The factor $1/2$ comes from
integrating over the half-space.

\subsection{Summary of the four kernel/surface combinations}

\begin{table}[ht]
\centering
\renewcommand{\arraystretch}{1.8}
\begin{tabular}{|l|l|c|}
\toprule
Kernel & Surface & $\chis = \chispar = \chisper$ \\
\midrule
Gaussian scalar
  & Plane
  & $\dfrac{\Delta_0\pi^{3/2}\ell^3}{2}$ \\[8pt]
Yukawa scalar
  & Sphere $R_0$
  & $2\pi\Delta_0\ell^3$ \\[8pt]
Lorentz tensorial ($\beta$)
  & Plane
  & $\dfrac{\pi^2\Delta_0\ell^3(6+\beta)}{12}$ \\[8pt]
Local ($\ell\to 0$)
  & Any
  & $\chi_0/2$ \\
\bottomrule
\end{tabular}
\caption{Leading-order surface susceptibility $\chi^s$ for the four
  kernel/surface combinations treated in Section~\ref{sec:explicit}. In every
  case the tangential and normal surface susceptibilities coincide
  ($\chispar=\chisper=\chi^s$), which is a general consequence of the
  combined rotational and parity invariance of the bulk kernel.}
\label{tab:explicit_chis}
\end{table}

Two conclusions emerge:
\begin{enumerate}
  \item \textbf{Isotropic centrosymmetric kernels give isotropic surface response}:
    $\chispar = \chisper = \chis$ in all cases, regardless of whether the
    bulk kernel is scalar or tensorial.
  \item \textbf{Curvature corrections scale as $\mu_2/R$}: for the
    Yukawa kernel on the sphere they are of relative order
    $6\ell^2/R_0$, observable for $R_0\sim 10\ell$.
\end{enumerate}


\section{Generalized Maxwell boundary conditions}
\label{sec:CL}

We now derive the boundary conditions associated with the
Maxwell equations in the presence of the surface distributions
established in Section~\ref{sec:3D}.  We work in the harmonic
regime at frequency $\omega$ with the $e^{-i\omega t}$ convention,
and set $\mathbf{M} = 0$ (no magnetic polarization).

\begin{remark}
  Throughout this section we retain the notation $\chispar$ and $\chisper$
  for the tangential and normal surface susceptibilities in order to
  facilitate comparison with the Feibelman literature, where
  $d_\parallel \neq d_\perp$ in general.  Within the present model,
  however, Section~\ref{sec:distrib3D} and Appendix~\ref{app:chi_derivation}
  establish that $\chispar = \chisper = \chis$ for any centrosymmetric
  isotropic bulk kernel.  All results below specialise to this case by
  setting $\chispar = \chisper$.
\end{remark}

\subsection{Maxwell equations in the distributional sense}

The macroscopic Maxwell equations in matter read:
\begin{align}
  \nabla\times\bE &= i\omega\mu_0\bH,
  \label{eq:MaxFar}\\
  \nabla\times\bH &= -i\omega\bD,
  \label{eq:MaxAmp}\\
  \nabla\cdot\bD  &= 0,
  \label{eq:MaxGauss}\\
  \nabla\cdot\bH  &= 0,
  \label{eq:MaxGaussH}
\end{align}
with $\bD = \varepsilon_0\bE + \bP$.  When the fields and the
polarization contain singular distributions supported on $\dOmega$,
these equations must be interpreted in the distributional sense.
We recall the standard distributional differentiation rules for a
field $\mathbf{F}$ discontinuous across $\dOmega$~\cite{Schwartz1966}:
\begin{align}
  \nabla\cdot\mathbf{F}
  &= \left\{\nabla\cdot\mathbf{F}\right\}
     + \jump{\mathbf{F}\cdot\bn}\,\delta_{\dOmega},
  \label{eq:div_distrib}\\
  \nabla\times\mathbf{F}
  &= \left\{\nabla\times\mathbf{F}\right\}
     + \bn\times\jump{\mathbf{F}}\,\delta_{\dOmega},
  \label{eq:rot_distrib}\\
  \nabla f
  &= \left\{\nabla f\right\}
     + \jump{f}\,\bn\,\delta_{\dOmega},
  \label{eq:grad_distrib}
\end{align}
where $\left\{\,\cdot\,\right\}$ denotes the regular part of the
distribution (identified with a locally integrable function away from
$\dOmega$) and $\jump{\mathbf{F}} := \mathbf{F}^+ - \mathbf{F}^-$
denotes the jump of $\mathbf{F}$ across $\dOmega$, with $\mathbf{F}^+$
the limit from $\Omega^c$ and $\mathbf{F}^-$ the limit from $\Omega$.

For a surface distribution $\mathbf{Q}^s\delta_{\dOmega}$:
\begin{align}
  \nabla\cdot(\mathbf{Q}^s\delta_{\dOmega})
  &= \nabla_s\cdot\mathbf{Q}^s_\parallel\,\delta_{\dOmega}
     - Q^s_\perp\,\partial_n\delta_{\dOmega},
  \label{eq:div_surf}\\
  \nabla\times(\mathbf{Q}^s\delta_{\dOmega})
  &= \nabla_s\times\mathbf{Q}^s_\parallel\,\delta_{\dOmega}
     + \bn\times\mathbf{Q}^s\,\partial_n\delta_{\dOmega},
  \label{eq:rot_surf}
\end{align}
where $\nabla_s$ is the surface gradient operator (Laplace--Beltrami),
$\mathbf{Q}^s_\parallel$ the tangential component, and
$Q^s_\perp = \mathbf{Q}^s\cdot\bn$ the normal component.

\subsection{Structure of the total polarization}

From the distributional expansion established in
Section~\ref{sec:distrib3D}, the total polarization splits into two terms~:
\begin{equation}\label{eq:P_total}
  \bP(\br) = \bP^{\Omega}(\br) + \bP^{\partial \Omega}(\br),
\end{equation}
where $\bP^{\Omega}$ is the regular bulk polarization and $\bP^{\partial \Omega}$
the interfacial contribution.  At the order retained here
(leading surface susceptibilities plus curvature corrections):
\begin{equation}\label{eq:deltaP_full}
\bP^{\partial \Omega}
  = -\varepsilon_0
  \Bigl[
    \mathcal{S}^{(0)}_{ij}\,E_j^+\,\delta_{\dOmega}
    + \bigl(\nu^\parallel H\, P^\parallel_{ij}
            + \nu^\perp  H\, P^\perp_{ij}\bigr)E_j^+\,\delta_{\dOmega}
    - \bigl(\nu^\parallel P^\parallel_{ij}
            + \nu^\perp  P^\perp_{ij}\bigr)E_j^+\,\partial_n\delta_{\dOmega}
  \Bigr]
  + O(\ell^3),
\end{equation}
where we recall $\mathcal{S}^{(0)}_{ij} =
\chispar P^\parallel_{ij} + \chisper P^\perp_{ij}$.

We separate~\eqref{eq:deltaP_full} into tangential and normal parts:
\begin{align}
  P^{\parallel,\partial \Omega}_i
  &= -\varepsilon_0
  \bigl[
    \chispar + \nu^\parallel H
  \bigr]
  E^+_{i,\parallel}\,\delta_{\dOmega}
  + \varepsilon_0\,\nu^\parallel\,
  E^+_{i,\parallel}\,\partial_n\delta_{\dOmega}
  + O(\ell^3),
  \label{eq:deltaPpar}\\[6pt]
  P^{\perp,\partial \Omega}
  &= -\varepsilon_0
  \bigl[
    \chisper + \nu^\perp H
  \bigr]
  E^+_\perp\,\delta_{\dOmega}
  + \varepsilon_0\,\nu^\perp\,
  E^+_\perp\,\partial_n\delta_{\dOmega}
  + O(\ell^3).
  \label{eq:deltaPperp}
\end{align}

\subsection{Identification of boundary conditions}
\label{sec:BC_step1}

We insert~\eqref{eq:deltaP_full} into the Maxwell
equations~\eqref{eq:MaxFar}--\eqref{eq:MaxGaussH} and identify the
coefficients of $\delta_{\dOmega}$ and $\partial_n\delta_{\dOmega}$.

\subsubsection{Faraday's law~(See \eqref{eq:MaxFar})}

In the absence of magnetic polarization, $\nabla\times\bE$ contains no
singularity contributed by $\delta\bP$.  Collecting the coefficient of
$\delta_{\dOmega}$ in~\eqref{eq:MaxFar} via~\eqref{eq:rot_distrib}
yields the classical tangential continuity condition:
\begin{equation}\label{eq:BC_Far}
\bn\times\jump{\bE} = \mathbf{0},
\end{equation}
i.e., $\bE^+_\parallel = \bE^-_\parallel$.  This condition is not
modified by the surface susceptibilities.

\subsubsection{Amp\`{e}re's law~(See \eqref{eq:MaxAmp})}

The surface part of $\bD = \varepsilon_0\bE + \bP^\Omega + \bP^{\partial \Omega}$
contributes to $\nabla\times\bH = -i\omega\bD$ via~\eqref{eq:rot_surf}.
Collecting the coefficient of $\delta_{\dOmega}$ gives:
\begin{equation}\label{eq:BC_Amp}
    \bn\times\jump{\bH}
    = -i\omega\varepsilon_0
    \Bigl[
      \bigl(\chispar + \nu^\parallel H\bigr)\bE^+_\parallel
      + \nu^\parallel\,\nabla_s\times\bE^+_\parallel
    \Bigr]
    + O(\ell^3).
\end{equation}
Here $\nabla_s\times\bE^+_\parallel$ is the surface curl of the
tangential field, which is a vector normal to $\dOmega$.  Collecting the coefficient of
$\partial_n\delta_{\dOmega}$ yields an additional condition on the
mean of the normal derivatives:
\begin{equation}\label{eq:BC_Amp_dn}
  \bn\times\avg{\partial_n\bH}
  = i\omega\varepsilon_0\,\nu^\parallel\,\bE^+_\parallel + O(\ell^3),
\end{equation}
where $\avg{\partial_n\bH} :=
(\partial_n\bH^+ + \partial_n\bH^-)/2$.

\subsubsection{Electric Gauss law~(See \eqref{eq:MaxGauss})}

Applying~\eqref{eq:div_distrib} and~\eqref{eq:div_surf} to
$\bD = \varepsilon_0\bE + \bP^\Omega + \delta\bP$, and collecting the
coefficient of $\delta_{\dOmega}$:
\begin{equation}\label{eq:BC_Gauss}
    \jump{D_\perp}
    = -\varepsilon_0
    \Bigl[
      \bigl(\chisper + \nu^\perp H\bigr)E^+_\perp
      + \nabla_s\cdot\bigl(\chispar\bE^+_\parallel\bigr)
    \Bigr]
    + O(\ell^3).
\end{equation}
The coefficient of $\partial_n\delta_{\dOmega}$ gives:
\begin{equation}\label{eq:BC_Gauss_dn}
  \avg{D_\perp} = \varepsilon_0\,\nu^\perp\,E^+_\perp + O(\ell^3).
\end{equation}

\subsubsection{Magnetic Gauss law~(See \eqref{eq:MaxGaussH})}

In the absence of magnetic polarization, the condition is classical:
\begin{equation}\label{eq:BC_GaussH}
\jump{H_\perp} = 0.
\end{equation}

\subsubsection{Intermediate summary}

\begin{table}[ht]
\centering
\renewcommand{\arraystretch}{2.2}
\begin{tabular}{|l|l|l|}
\toprule
Equation & Leading condition & Curvature correction ($\ell^2$) \\
\midrule
Faraday
  & $\bn\times\jump{\bE} = \mathbf{0}$
  & unchanged \\
Amp\`{e}re
  & $\bn\times\jump{\bH}
     = -i\omega\varepsilon_0\chispar\bE^+_\parallel$
  & $+\,\nu^\parallel
       (H\bE^+_\parallel + \nabla_s\times\bE^+_\parallel)$ \\
Gauss (elec.)
  & $\jump{D_\perp}
     = -\varepsilon_0(\chisper E^+_\perp
       + \nabla_s\cdot\chispar\bE^+_\parallel)$
  & $-\,\varepsilon_0\nu^\perp H E^+_\perp$ \\
Gauss (mag.)
  & $\jump{H_\perp} = 0$
  & unchanged \\
\bottomrule
\end{tabular}
\caption{Generalised Maxwell boundary conditions for the linear nonlocal
  response at leading order, together with the explicit curvature
  corrections of order~$\ell^2$ (proportional to the mean curvature~$H$).
  Faraday and the magnetic Gauss law are unchanged by the surface
  susceptibilities; the modifications affect only Amp\`ere and the electric
  Gauss law.}
\label{tab:BC_summary}
\end{table}

\subsection{Resolution of the implicit equation on $E^+_\perp$}
\label{sec:BC_step2}

Condition~\eqref{eq:BC_Gauss} is \emph{implicit} in $E^+_\perp$: the
right-hand side contains $E^+_\perp$, while the left-hand side
$\jump{D_\perp} = \varepsilon_{\mathrm{ext}}E^+_\perp
- \varepsilon_{\mathrm{int}}E^-_\perp$ also depends on it (here
$\varepsilon_{\mathrm{ext}}$ and $\varepsilon_{\mathrm{int}}$ are the
bulk permittivities of $\Omega^c$ and $\Omega$, respectively).
At leading order~\eqref{eq:BC_Gauss} reads:
\begin{equation}\label{eq:implicit}
  \varepsilon_{\mathrm{ext}}\,E^+_\perp
  - \varepsilon_{\mathrm{int}}\,E^-_\perp
  = -\varepsilon_0\chisper\,E^+_\perp
  - \varepsilon_0\nabla_s\cdot\!\bigl(\chispar\bE^+_\parallel\bigr).
\end{equation}
Collecting the $E^+_\perp$ terms and introducing the
\emph{renormalized exterior permittivity}
\begin{equation}\label{eq:eps_renorm}
  \tilde\varepsilon_{\mathrm{ext}}
  := \varepsilon_{\mathrm{ext}} + \varepsilon_0\chisper,
\end{equation}
one obtains explicitly:
\begin{equation}\label{eq:Eperp_explicit}
    E^+_\perp
    = \frac{\varepsilon_{\mathrm{int}}}
           {\tilde\varepsilon_{\mathrm{ext}}}\,E^-_\perp
    - \frac{\varepsilon_0}{\tilde\varepsilon_{\mathrm{ext}}}
      \nabla_s\cdot\!\bigl(\chispar\bE^+_\parallel\bigr).
\end{equation}

\begin{remark}[Physical meaning of the renormalization]
  The surface susceptibility $\chisper$ renormalizes the effective
  permittivity on the exterior side: the transition layer behaves as
  if the exterior medium had permittivity
  $\tilde\varepsilon_{\mathrm{ext}} = \varepsilon_{\mathrm{ext}} +
  \varepsilon_0\chisper$.  For a denser-than-vacuum interfacial layer
  ($\chisper > 0$), $\tilde\varepsilon_{\mathrm{ext}} >
  \varepsilon_{\mathrm{ext}}$: the layer enhances the effective
  permittivity seen by the exterior field.
\end{remark}

\subsection{Explicit boundary conditions}

Substituting~\eqref{eq:Eperp_explicit} into the conditions of
Section~\ref{sec:BC_step1} and using $\bE^+_\parallel =
\bE^-_\parallel$ (Faraday), we obtain conditions involving only the
interior field $\bE^-$ and the exterior tangential field, both
well-defined quantities.

\paragraph{Jump of tangential $\bH$.}
\begin{equation}\label{eq:BC_H_explicit}
    \bn\times\jump{\bH}
    = -i\omega\varepsilon_0\,\chispar\,\bE^-_\parallel
    - i\omega\varepsilon_0\,\nu^\parallel
      \Bigl(
        H\,\bE^-_\parallel
        + \nabla_s\times\bE^-_\parallel
      \Bigr)
    + O(\ell^3).
\end{equation}

\paragraph{Jump of normal $D$.}
\begin{equation}\label{eq:BC_D_explicit}
    \varepsilon_{\mathrm{ext}}\,E^+_\perp
    - \varepsilon_{\mathrm{int}}\,E^-_\perp
    = -\frac{\varepsilon_0\,\varepsilon_{\mathrm{int}}\,\chisper}
            {\tilde\varepsilon_{\mathrm{ext}}}\,E^-_\perp
    - \frac{\varepsilon_0}{\tilde\varepsilon_{\mathrm{ext}}}
      \bigl(1 + \nu^\perp H\bigr)
      \nabla_s\cdot\!\bigl(\chispar\bE^-_\parallel\bigr)
    + O(\ell^3).
\end{equation}

\begin{remark}[Universality and experimental access]
  The two conditions~\eqref{eq:BC_H_explicit}--\eqref{eq:BC_D_explicit}
  are universal: the entire microscopic complexity of the nonlocal kernel
  is encoded in just four effective scalars, $\chispar$, $\chisper$,
  $\nu^\parallel$, $\nu^\perp$.  As shown in
  Section~\ref{sec:Feibelman_d} below, $\chispar$ and $\chisper$ are
  proportional to the Feibelman $d$-parameters, which are routinely
  extracted from \emph{flat} interfaces by spectroscopic ellipsometry
  (angular and frequency dependence of the Fresnel coefficients $r_p$,
  $r_s$) or from the dispersion of surface plasmons measured by
  attenuated total reflection or electron energy-loss spectroscopy.
  The curvature parameters $\nu^\parallel$ and $\nu^\perp$ appear
  multiplied by the mean curvature $H$ and are accessible by comparing
  interfaces of different curvatures: for spherical nanoparticles the
  plasmonic Mie resonances are shifted from their classical positions by
  an amount proportional to $d_\perp/R$, and measuring this shift as a
  function of radius $R$ in single-particle dark-field spectroscopy
  isolates $\nu^\perp$ (and similarly $\nu^\parallel$ from non-spherical
  geometries).  All four parameters can therefore be determined
  experimentally without any knowledge of the underlying nonlocal kernel.
\end{remark}

\subsection{Connection with the Feibelman $d$-parameters}
\label{sec:Feibelman_d}

The Feibelman $d$-parameters are defined as~\cite{Feibelman1982}:
\begin{equation}\label{eq:Feibelman_def}
  d_\perp(\omega)
  := \frac{\int z\,\delta\varepsilon(z,\omega)\,E_\perp(z)\,\diff z}
          {\jump{\varepsilon E_\perp}},
  \qquad
  d_\parallel(\omega)
  := \frac{\int z\,\partial_z E_\parallel(z)\,\diff z}
          {\jump{E_\parallel}},
\end{equation}
where $\delta\varepsilon(z)$ is the excess permittivity in the
transition layer.  In the present formalism these parameters emerge as:
\begin{equation}\label{eq:Feibelman_link}
  d_\perp \sim
  \frac{\varepsilon_0\,\chisper}{\jump\varepsilon},
  \qquad
  d_\parallel \sim
  \frac{\varepsilon_0\,\chispar}{\jump\varepsilon},
\end{equation}
where $\jump\varepsilon = \varepsilon_{\mathrm{ext}} -
\varepsilon_{\mathrm{int}}$.  The curvature parameters $\nu^\parallel$
and $\nu^\perp$ constitute the natural extension of the Feibelman
$d$-parameters to curved interfaces:
\begin{equation}\label{eq:Feibelman_curved}
  d_\perp^{\mathrm{curv}}
  \sim
  \frac{\varepsilon_0(\chisper + \nu^\perp H)}{\jump\varepsilon},
  \qquad
  d_\parallel^{\mathrm{curv}}
  \sim
  \frac{\varepsilon_0(\chispar + \nu^\parallel H)}{\jump\varepsilon}.
\end{equation}
The flat-interface Feibelman parameters are thus the zeroth-order
members of a hierarchy of surface coefficients, each controlled by a
successive moment of the nonlocal kernel.

\subsection{Special cases}

\subsubsection{Planar interface}

For $\dOmega = \{z=0\}$ one has $H=K=0$ and, by Faraday's law in the
bulk, $\nabla_s\times\bE^-_\parallel = 0$.
Conditions~\eqref{eq:BC_H_explicit}--\eqref{eq:BC_D_explicit}
reduce to:
\begin{align}
  \bn\times\jump{\bH}
  &= -i\omega\varepsilon_0\,\chispar\,\bE^-_\parallel,
  \label{eq:BC_plane_H}\\[6pt]
  \jump{D_\perp}
  &= -\frac{\varepsilon_0\,\varepsilon_{\mathrm{int}}\,\chisper}
           {\tilde\varepsilon_{\mathrm{ext}}}\,E^-_\perp
     - \frac{\varepsilon_0}{\tilde\varepsilon_{\mathrm{ext}}}
       \nabla_s\cdot\!\bigl(\chispar\bE^-_\parallel\bigr).
  \label{eq:BC_plane_D}
\end{align}
For $\chispar = \chisper = 0$ one recovers $\bn\times\jump{\bH}=0$
and $\jump{D_\perp}=0$: the classical Fresnel conditions.

\subsubsection{Spherical interface}

For the sphere of radius $R_0$ ($H = 2/R_0$, $\bn = \hat{\mathbf{r}}$):
\begin{align}
  \bn\times\jump{\bH}
  &= -i\omega\varepsilon_0
     \!\left(\chispar + \frac{2\nu^\parallel}{R_0}\right)
     \bE^-_\parallel
     - i\omega\varepsilon_0\,\nu^\parallel\,
       \nabla_s\times\bE^-_\parallel,
  \label{eq:BC_sphere_H}\\[6pt]
  \jump{D_\perp}
  &= -\frac{\varepsilon_0\,\varepsilon_{\mathrm{int}}}
           {\tilde\varepsilon_{\mathrm{ext}}}
     \!\left(\chisper + \frac{2\nu^\perp}{R_0}\right)E^-_\perp
     - \frac{\varepsilon_0}{\tilde\varepsilon_{\mathrm{ext}}}
       \!\left(1 + \frac{2\nu^\perp}{R_0}\right)
       \nabla_s\cdot\!\bigl(\chispar\bE^-_\parallel\bigr).
  \label{eq:BC_sphere_D}
\end{align}
The terms proportional to $1/R_0$ are the nonlocal Mie corrections:
they shift the scattering resonances of a plasmonic nanosphere with
respect to the planar case, and become significant for
$R_0\lesssim 10\ell$.

\begin{remark}[Limiting cases]
  Setting $\chispar = \chisper = \nu^\parallel = \nu^\perp = 0$
  in~\eqref{eq:BC_sphere_H}--\eqref{eq:BC_sphere_D} recovers the
  classical Mie boundary conditions.  Setting $R_0\to\infty$
  recovers the planar
  conditions~\eqref{eq:BC_plane_H}--\eqref{eq:BC_plane_D}.
\end{remark}


\section{Numerical examples}
\label{sec:numerics}

\subsection{1D benchmarks: planar interface and thin film}
\label{sec:numerics_1D}

Before treating the curved geometry, we validate the framework on two
planar configurations that admit closed-form solutions and connect
directly to standard thin-film optics.

\subsubsection{Planar interface: modified Fresnel coefficients}
\label{sec:fresnel_interface}

We consider a planar interface at $z=0$ between vacuum ($n_1 = 1$) and
a nonlocal medium of bulk index $n_2 = 1.5$, illuminated by a TE-polarised
plane wave at angle of incidence $\theta_i$.  As shown in
Appendix~\ref{app:fresnel}, the modified boundary condition
\eqref{eq:BC_plane_H} yields
\begin{equation}\label{eq:rTE_main}
  r_{\mathrm{TE}}
  = \frac{n_1\cos\theta_i - n_2\cos\theta_t + i(\omega/c)\,\chispar}
         {n_1\cos\theta_i + n_2\cos\theta_t - i(\omega/c)\,\chispar},
\end{equation}
with the transmission amplitude $t_{\mathrm{TE}} = 1 + r_{\mathrm{TE}}$;
at $\chispar = 0$ the standard Fresnel coefficient is recovered.

Two remarks put \eqref{eq:rTE_main} in perspective:

\begin{enumerate}
  \item \textbf{Reflectance correction is second order.}
    Writing $r_{\mathrm{TE}} = r_0 (1 + 2i\gamma / (r_0 D))$ with
    $r_0 = (n_1 c_i - n_2 c_t)/(n_1 c_i + n_2 c_t)$,
    $D = n_1 c_i + n_2 c_t$
    ($c_j = \cos\theta_j$), and $\gamma = k_0\chispar$, one finds
    $|r_{\mathrm{TE}}|^2 = r_0^2 + O(\gamma^2)$:
    the \emph{intensity} reflectance is unaffected at first order in $\chispar$.

  \item \textbf{Phase correction is first order.}
    In contrast, $\arg r_{\mathrm{TE}}$ acquires a correction of order
    $\gamma \sim k_0\chispar \sim \ell/\lambda$, measurable by ellipsometry
    even when $\ell \ll \lambda$.
\end{enumerate}

\begin{figure}[h!]
  \centering
  \includegraphics[width=\linewidth]{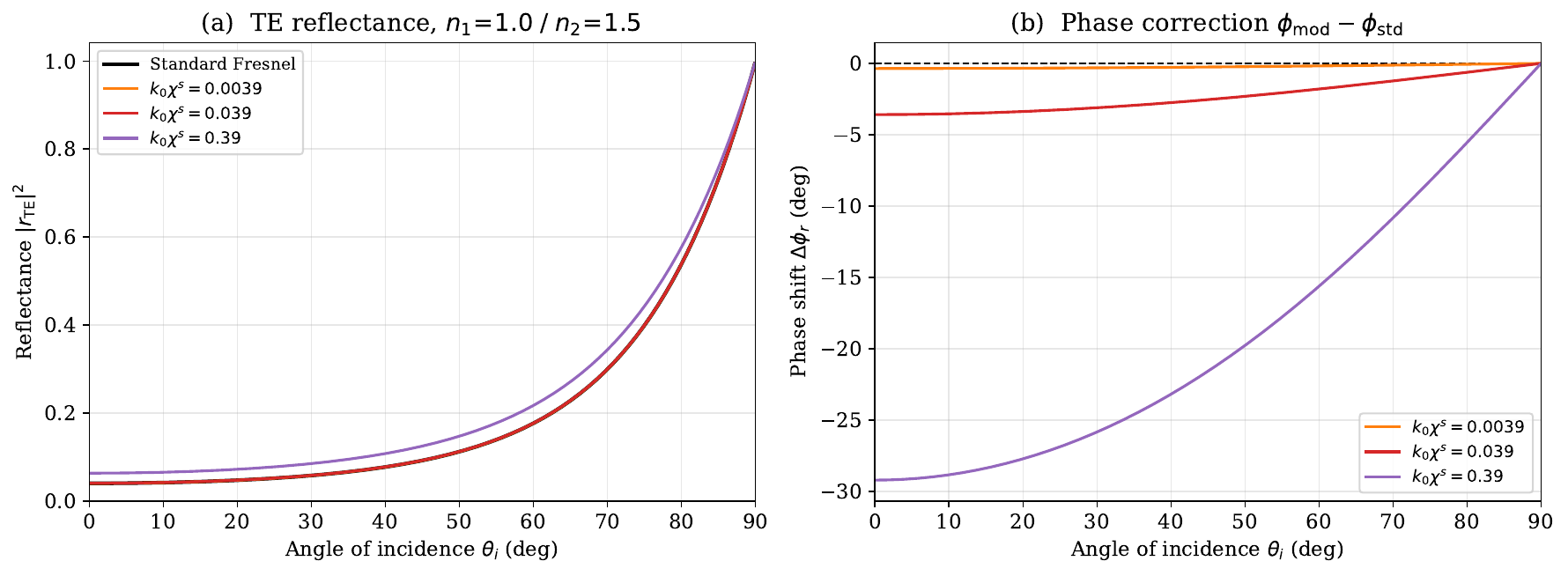}
  \caption{Modified Fresnel coefficients for the TE polarisation at a
           planar $\text{vacuum}/n_2$ interface ($n_2 = 1.5$).
           \textbf{(a)} Reflectance $|r_{\mathrm{TE}}|^2$ vs angle of incidence.
           \textbf{(b)} Phase shift $\Delta\phi_r = \arg r_{\mathrm{mod}} - \arg r_{\mathrm{std}}$
           (in degrees).  Three curves correspond to $k_0\chispar = 3.9\times10^{-3}$
           (physical value for $\ell/\lambda = 10^{-3}$), $3.9\times10^{-2}$, and $0.39$.
           For TE polarisation the reflectance has no zero crossing (no Brewster
           angle); the phase correction is smooth and monotone across all angles.}
  \label{fig:fresnel_interface}
\end{figure}

Figure~\ref{fig:fresnel_interface} illustrates both effects for three
values of $k_0\chispar$ spanning the range $[4\times10^{-3},\,0.4]$.
In panel~(a), the reflectance curves are nearly indistinguishable from
the standard result at the physical value ($k_0\chispar \approx 4\times10^{-3}$)
and only become visible for $k_0\chispar \sim 0.1$--$0.4$.  In contrast,
panel~(b) shows that the reflection phase is shifted by a quantity
proportional to $k_0\chispar$ across the entire angular range, including
a smooth, angle-dependent phase correction that is measurable by ellipsometry
even when $\ell \ll \lambda$.
This phase sensitivity is precisely what is exploited in precision
ellipsometry to detect sub-nanometre interface layers.

\subsubsection{Thin film: modified Fabry--P\'{e}rot transmittance}
\label{sec:fresnel_film}

\begin{figure}[h!]
  \centering

\begin{tikzpicture}[>=Stealth, thick, font=\small]

  \def\figH{3.2}      
  \def\vacW{2.0}      
  \def\slabW{3.5}     
  \def\layerW{0.35}   

  \pgfmathsetmacro{\xA}{0}                              
  \pgfmathsetmacro{\xB}{\vacW}                          
  \pgfmathsetmacro{\xC}{\vacW+\layerW}                  
  \pgfmathsetmacro{\xD}{\vacW+\slabW+\layerW}           
  \pgfmathsetmacro{\xE}{\vacW+\slabW+2*\layerW}         
  \pgfmathsetmacro{\xF}{\vacW+\slabW+2*\layerW+\vacW}   

  \fill[blue!10]   (\xA,0) rectangle (\xB,\figH);   
  \fill[orange!25] (\xB,0) rectangle (\xE,\figH);   
  \fill[orange!55] (\xB,0) rectangle (\xC,\figH);   
  \fill[orange!55] (\xD,0) rectangle (\xE,\figH);   
  \fill[blue!10]   (\xE,0) rectangle (\xF,\figH);   

  \draw (\xA,0) rectangle (\xF,\figH);  
  \draw (\xB,0) -- (\xB,\figH);         
  \draw (\xE,0) -- (\xE,\figH);         

  \node[align=center] at ({(\xA+\xB)/2}, {\figH/2+0.22}) {vacuum};
  \node                at ({(\xA+\xB)/2}, {\figH/2-0.22}) {$n=1$};

  \node[align=center] at ({(\xC+\xD)/2}, {\figH/2+0.22}) {glass};
  \node                at ({(\xC+\xD)/2}, {\figH/2-0.22}) {$n_f = 1.5$};

  \node[align=center] at ({(\xE+\xF)/2}, {\figH/2+0.22}) {vacuum};
  \node                at ({(\xE+\xF)/2}, {\figH/2-0.22}) {$n=1$};

  \draw[->, decorate,
        decoration={snake, amplitude=2.5pt, segment length=9pt,
                    post length=5pt}]
    (\xA+0.15, \figH*0.72) -- (\xB-0.05, \figH*0.72)
    node[pos=0.45, above=2pt] {$E_i$};

  \draw[<-, decorate,
        decoration={snake, amplitude=2.5pt, segment length=9pt,
                    post length=5pt}]
    (\xA+0.15, \figH*0.35) -- (\xB-0.05, \figH*0.35)
    node[pos=0.45, below=2pt] {$E_r$};

  \draw[->, dashed, gray!70, thin]
    (\xC+0.08, \figH*0.70) -- (\xD-0.08, \figH*0.70);
  \draw[->, dashed, gray!70, thin]
    (\xD-0.08, \figH*0.54) -- (\xC+0.08, \figH*0.54);
  \draw[->, dashed, gray!70, thin]
    (\xC+0.08, \figH*0.38) -- (\xD-0.08, \figH*0.38);

  \draw[->, decorate,
        decoration={snake, amplitude=2.5pt, segment length=9pt,
                    post length=5pt}]
    (\xE+0.05, \figH*0.75) -- (\xF-0.15, \figH*0.75)
    node[pos=0.55, above=2pt] {$E_t$};

  \node[above, font=\footnotesize] at ({(\xB+\xC)/2}, \figH)
    {$\chispar$};
  \node[above, font=\footnotesize] at ({(\xD+\xE)/2}, \figH)
    {$\chispar$};

  \draw[|<->|, thin] (\xB, -0.75) -- (\xE, -0.75)
    node[midway, below] {$d$};
  \draw[thin, gray!60, dashed] (\xB,0) -- (\xB,-0.75);
  \draw[thin, gray!60, dashed] (\xE,0) -- (\xE,-0.75);

  \draw[|<->|, thin] (\xB, -0.20) -- (\xC, -0.20)
    node[midway, below, font=\footnotesize] {${\sim}\,\ell$};

\end{tikzpicture}
  \caption{Thin-film Fabry--P\'erot geometry.  A glass slab of index
    $n_f = 1.5$ and thickness $d$ is surrounded by vacuum ($n=1$) on
    both sides.  Each interface carries a nonlocal surface layer of
    characteristic thickness $\sim\ell$ (darker shading), encoded by
    the surface susceptibility $\chispar$.  A plane wave at normal
    incidence ($E_i$) generates a reflected field ($E_r$), multiple
    internal round-trips (dashed), and a transmitted field ($E_t$).}
  \label{fig:fabry_perot_sketch}
\end{figure}
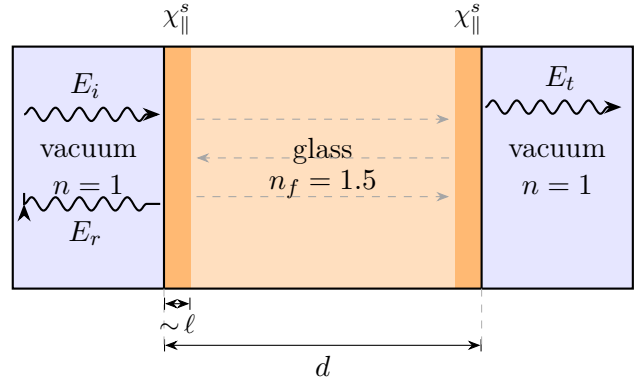

We now consider a free-standing glass slab of index $n_f = n_2 = 1.5$
and thickness $d$, surrounded by vacuum on both sides
(Fig.~\ref{fig:fabry_perot_sketch}).  Each of the two
interfaces carries the same surface susceptibility $\chispar = (n_2^2-1)\ell/2$
from the consistency relation~\eqref{eq:chis_consistency}.  Using the
transfer-matrix formalism with the modified interface
matrix~\eqref{eq:P_limit}, the transmittance can be evaluated exactly
for any $d$ and $\ell$.

The practical significance of this setting is that the thin-film
community routinely introduces \emph{ad hoc} corrections to the
standard Fabry--P\'{e}rot formula when fitting ellipsometric or
transmittance data for ultrathin films: empirical ``dead layers'',
effective-medium adjustments, or phenomenological interface roughness
parameters are used as fitting knobs.  Our framework shows that these
corrections have a systematic origin: they are entirely encoded in
$\chispar$, which is fixed (without any free parameter) by the first
moment of the nonlocal kernel.

\begin{figure}[h!]
  \centering
  \includegraphics[width=\linewidth]{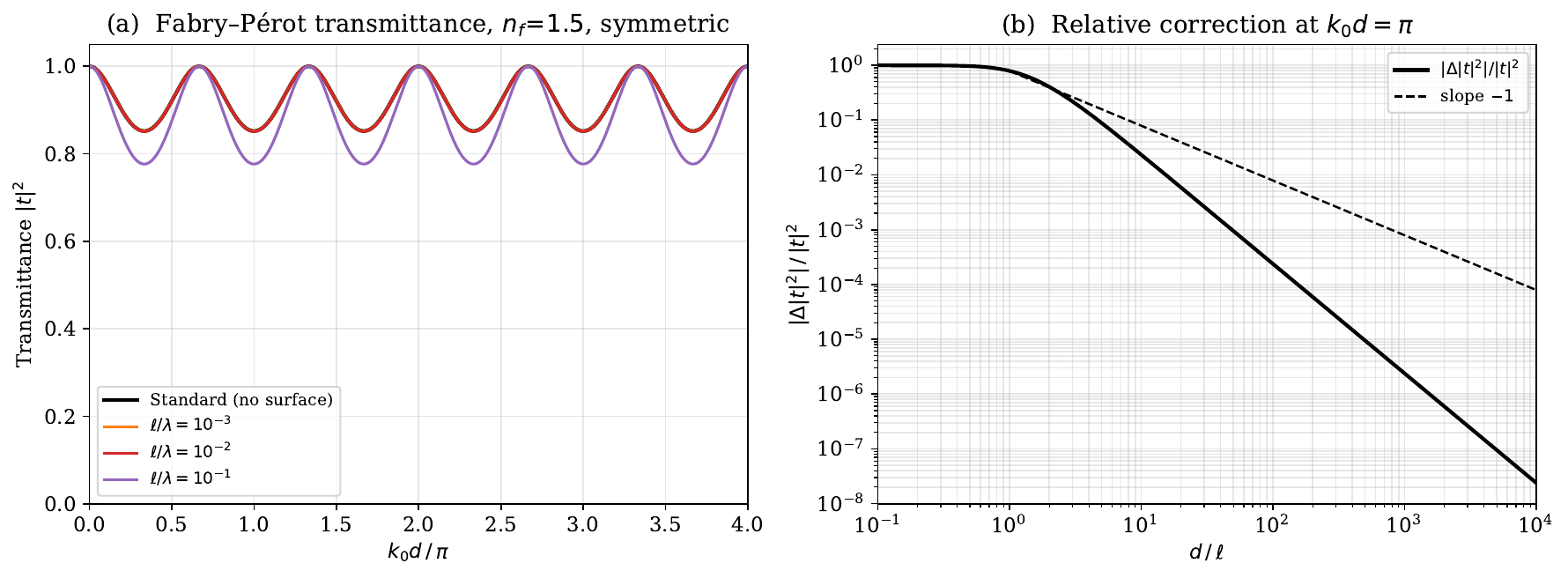}
  \caption{Modified Fabry--P\'{e}rot transmittance for a free-standing
           glass slab ($n_f = 1.5$, normal incidence).
           \textbf{(a)} $|t|^2$ vs $k_0d/\pi$ for three values of $\ell/\lambda$;
           the standard result (no surface susceptibility) is shown in black.
           \textbf{(b)} Relative correction $|\Delta|t|^2|/|t|^2$ at
           $k_0 d = \pi$ vs $d/\ell$ in log-log scale; the dashed line has
           slope $-1$, confirming the $\ell/d$ scaling for $d \gg \ell$.}
  \label{fig:fresnel_film}
\end{figure}

Figure~\ref{fig:fresnel_film} quantifies this claim:
\begin{itemize}
  \item \textbf{Panel~(a)} shows $|t|^2$ versus the reduced film
    thickness $k_0 d/\pi$ for three values of $\ell/\lambda$.
    At $\ell/\lambda = 10^{-3}$ (physical regime) the correction is
    invisible on the plot; at $\ell/\lambda = 10^{-2}$ the Fabry--P\'{e}rot
    fringes begin to shift; at $\ell/\lambda = 10^{-1}$ the deviation
    from standard optics is $O(1)$.
  \item \textbf{Panel~(b)} plots the relative correction
    $|\Delta|t|^2|/|t|^2$ at the first Fabry--P\'{e}rot resonance
    ($k_0 d = \pi$) as a function of $d/\ell$.  The log-log slope
    is $-1$, confirming the scaling law
    \begin{equation}\label{eq:film_scaling}
      \frac{|\Delta|t|^2|}{|t|^2} \;\sim\; \frac{\ell}{d}
      \qquad (d \gg \ell).
    \end{equation}
    \emph{For $d \lesssim \ell$ the correction saturates to an $O(1)$ value:
    the standard thin-film formula then fails completely, and the
    surface susceptibility becomes the leading contribution.}
\end{itemize}

\subsection{2D Mie scattering with nonlocal surface susceptibilities}
\label{sec:numerics_2D}

\subsubsection{Setup}

\begin{figure}[h!]
  \centering

\begin{tikzpicture}[>=Stealth, thick, font=\small]

  \def\R{1.6}        
  \def\lT{0.20}      
  \def\cx{4.2}       
  \def\cy{2.8}       
  \def\figW{9.8}     
  \def\figH{5.6}     

  \pgfmathsetmacro{\RlT}{\R - \lT}              
  \pgfmathsetmacro{\xLeft}{\cx - \R}            
  \pgfmathsetmacro{\xRight}{\cx + \R}           
  \pgfmathsetmacro{\EsX}{\cx + (\R+1.15)*cos(45)}
  \pgfmathsetmacro{\EsY}{\cy + (\R+1.15)*sin(45)}
  \pgfmathsetmacro{\RadLabelX}{\cx + 0.55*\R*cos(135)}
  \pgfmathsetmacro{\RadLabelY}{\cy + 0.55*\R*sin(135)}
  \pgfmathsetmacro{\RadEndX}{\cx + (\R - \lT*0.5)*cos(135)}
  \pgfmathsetmacro{\RadEndY}{\cy + (\R - \lT*0.5)*sin(135)}
  \pgfmathsetmacro{\ChiArrowTipX}{\cx + (\R - \lT*0.5)*cos(20)}
  \pgfmathsetmacro{\ChiArrowTipY}{\cy + (\R - \lT*0.5)*sin(20)}
  \pgfmathsetmacro{\ChiLabelX}{\cx + \R + 0.55}
  \pgfmathsetmacro{\ChiLabelY}{\cy + 0.9}

  \fill[blue!10] (0,0) rectangle (\figW,\figH);
  \fill[orange!60] (\cx,\cy) circle (\R);
  \fill[orange!30] (\cx,\cy) circle (\RlT);

  \draw (0,0) rectangle (\figW,\figH);
  \draw[thick] (\cx,\cy) circle (\R);
  \draw[thick] (\cx,\cy) circle (\RlT);

  \foreach \yy in {1.4, 2.1, 2.8, 3.5, 4.2} {
    \draw[->, decorate,
          decoration={snake, amplitude=2pt, segment length=9pt,
                      post length=4pt}, blue!60, semithick]
      (0.2, \yy) -- ({\xLeft - 0.12}, \yy);
  }
  \draw[->, thick, blue!70] (0.3, 0.55) -- (1.3, 0.55);
  \node[right, blue!70, font=\footnotesize] at (1.35, 0.55) {$\mathbf{k}_i$};
  \node[above, blue!70, font=\footnotesize] at (1.0, 4.35) {$E_i$};

  \pgfmathsetmacro{\Rsc}{\R + 0.95}
  \draw[dashed, gray!60, thin] (\cx,\cy) circle (\Rsc);
  \foreach \ang in {20, 70, 120, 170, 220, 270, 320} {
    \pgfmathsetmacro{\xs}{\cx + (\R+0.22)*cos(\ang)}
    \pgfmathsetmacro{\ys}{\cy + (\R+0.22)*sin(\ang)}
    \pgfmathsetmacro{\xe}{\cx + (\R+0.80)*cos(\ang)}
    \pgfmathsetmacro{\ye}{\cy + (\R+0.80)*sin(\ang)}
    \draw[->, gray!55, thin] (\xs,\ys) -- (\xe,\ye);
  }
  \node[font=\footnotesize, gray!70] at (\EsX, \EsY) {$E_s$};

  \draw[thin] (\cx,\cy) -- (\RadEndX,\RadEndY);
  \node[font=\footnotesize, above left=-2pt] at (\RadLabelX,\RadLabelY) {$R_0$};

  \draw[|<->|, thin]
    ({\xRight - \lT}, {\cy - 0.18}) -- ({\xRight}, {\cy - 0.18})
    node[midway, below, font=\footnotesize] {${\sim}\,\ell$};
  \draw[thin, gray!50, dashed]
    ({\xRight - \lT}, {\cy - \lT*0.5}) -- ({\xRight - \lT}, {\cy - 0.18});

  \draw[->, thin, orange!70!black]
    (\ChiLabelX, \ChiLabelY)
    to[out=180, in=30] (\ChiArrowTipX, \ChiArrowTipY);
  \node[right, font=\footnotesize, orange!70!black]
    at (\ChiLabelX, \ChiLabelY) {$\chispar,\;\chisper$};

  \node[align=center] at (\cx, {\cy+0.25}) {glass};
  \node[align=center] at (\cx, {\cy-0.25}) {$n_2 = 1.5$};
  \node[right, font=\footnotesize, blue!60] at (0.18, {\figH-0.38})
    {vacuum \; $n_1=1$};

  \draw[rounded corners=3pt, gray!40, fill=white, fill opacity=0.85]
    ({\figW-3.50}, 0.12) rectangle ({\figW-0.12}, 1.18);
  \node[align=left, font=\footnotesize] at ({\figW-1.85}, 0.65) {
    \textbf{TM}: $E_z\neq 0$, $H_z=0$\\[2pt]
    \textbf{TE}: $H_z\neq 0$, $E_z=0$
  };

\end{tikzpicture}
  \caption{2D Mie scattering geometry.  An infinite dielectric cylinder
    of radius $R_0$ and bulk index $n_2 = 1.5$ is embedded in vacuum
    ($n_1 = 1$).  A plane wave $E_i$ propagates in the $+x$ direction
    and generates a scattered field $E_s$.  The nonlocal surface layer
    of thickness $\sim\ell$ (darker annular ring) is characterised by
    $\chispar$ (TM polarisation, $E_z$) and both $\chispar$, $\chisper$
    (TE polarisation, $H_z$).  Curvature $H = 1/R_0$.}
  \label{fig:mie_2d_sketch}
\end{figure}
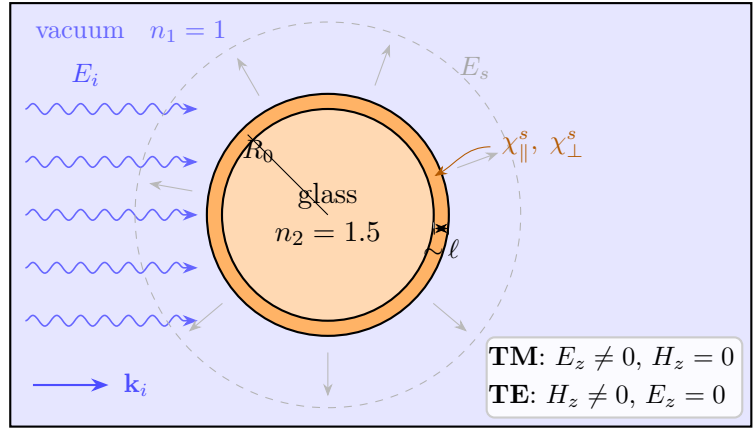

We illustrate the general boundary conditions of Section~\ref{sec:CL}
on a canonical geometry (Fig.~\ref{fig:mie_2d_sketch}) that admits a fully explicit solution: the
two-dimensional scattering of a plane wave by an infinite circular
dielectric cylinder of radius $R_0$ and bulk index $n_2$, surrounded
by vacuum ($n_1 = 1$).  This geometry has mean curvature $H = 1/R_0$
and Gaussian curvature $K = 0$, and its cylindrical symmetry
allows the scattered field to be expanded in a Fourier--Bessel series,
giving rise to modified Mie coefficients that can be computed analytically.

We treat both polarizations:
\begin{itemize}
  \item \textbf{TM polarization} ($\mathbf{E} = E_z\,\hat{\mathbf{z}}$,
    electric field parallel to the cylinder axis):
    the only non-zero component of $\mathbf{E}$ is tangential to the
    cylindrical surface, so the modified boundary condition involves
    $\chispar$ alone.
  \item \textbf{TE polarization} ($\mathbf{H} = H_z\,\hat{\mathbf{z}}$,
    magnetic field parallel to the cylinder axis):
    $\mathbf{E}$ lies in the $xy$-plane and has both a tangential
    component $E_\varphi$ and a normal component $E_r$; the two modified
    boundary conditions involve \emph{both} $\chispar$ and $\chisper$.
\end{itemize}

The surface susceptibilities are fixed by the consistency relation
derived in Section~\ref{sec:CL} for the exponential kernel
(Section~\ref{sec:expo}):
\begin{equation}\label{eq:chis_consistency}
  \chispar = \chisper = \frac{n_2^2 - 1}{2}\,\ell,
\end{equation}
where $\ell$ is the nonlocality range.  All numerical results use
$n_2 = 1.5$ and $\ell/\lambda = 10^{-3}$, so that $k_0\chispar \approx
3.9\times 10^{-3}$: the problem is in the perturbative nonlocal regime.

\subsubsection{Modified Mie coefficients}
\label{sec:mie_modified}

\paragraph{TM polarization.}

The incident plane wave $E_z^{\mathrm{inc}} = e^{ik_0 x}$ is decomposed
in cylindrical harmonics as $\sum_m i^m J_m(k_0 r)\,e^{im\varphi}$.
The scattered and interior fields have the same angular structure, with
coefficients $b_m$ (scattered, $H_m^{(1)}$ outside) and $c_m$ (interior,
$J_m$ inside).  The two boundary conditions at $r = R_0$ are:
\begin{align}
  &\text{(Faraday)}\quad
  [\![E_z]\!] = 0,
  \label{eq:BC_TM_1}\\
  &\text{(modified Amp\`{e}re)}\quad
  [\![\partial_r E_z]\!] = -k_0^2\,\chispar\,E_z\big|_{R_0}.
  \label{eq:BC_TM_2}
\end{align}
Inserting the Bessel expansions and solving the $2\times 2$ system,
one obtains:
\begin{equation}\label{eq:bm_TM}
b_m^{\mathrm{TM}} = -i^m\,
  \frac{k_0 J_m'(\rho_1) J_m(\rho_2)
       - k_2 J_m(\rho_1) J_m'(\rho_2)
       + k_0^2\chispar\, J_m(\rho_1) J_m(\rho_2)}
       {k_0 H_m^{(1)'}(\rho_1) J_m(\rho_2)
       - k_2 H_m^{(1)}(\rho_1) J_m'(\rho_2)
       + k_0^2\chispar\, H_m^{(1)}(\rho_1) J_m(\rho_2)},
\end{equation}
where $\rho_1 = k_0 R_0$, $\rho_2 = k_2 R_0$, and primes denote
derivatives with respect to the argument.  For $\chispar = 0$ this
reduces to the standard 2D TM Mie coefficient.

\begin{remark}
  The surface susceptibility $\chispar$ adds a term proportional to
  $J_m(\rho_1) J_m(\rho_2)$ in both numerator and denominator of
  $b_m$.  For $k_0\chispar \ll 1$, the correction is perturbative and
  shifts the Mie resonances (zeros of the denominator) by a complex
  amount of order $k_0\chispar$.
\end{remark}

\paragraph{TE polarization.}

For the TE case the field $H_z$ satisfies the same Helmholtz equation,
and the physical components $E_r = (m/\omega\varepsilon r)\,H_z$ and
$E_\varphi = (1/i\omega\varepsilon)\,\partial_r H_z$ are derived from it.
The three modified boundary conditions reduce, after elimination of the
interior coefficient $d_m$ via the continuity of $E_\varphi$, to a
single scalar equation per mode $m$:
\begin{align}
  &\text{(Faraday)}\quad
  n_2\bigl(i^m J_m'(\rho_1) + a_m H_m^{(1)'}(\rho_1)\bigr)
  = d_m J_m'(\rho_2),
  \label{eq:BC_TE_1}\\
  &\text{(modified Gauss + Amp\`{e}re)}\quad
  (1+\chisper)\,A_m - B_m
  = -i k_0\chispar\,\bigl(i^m J_m'(\rho_1) + a_m H_m^{(1)'}(\rho_1)\bigr),
  \label{eq:BC_TE_2}
\end{align}
where $A_m = i^m J_m(\rho_1) + a_m H_m^{(1)}(\rho_1)$ and
$B_m = d_m J_m(\rho_2)$.  The second condition is derived from the
distributional divergence identity applied to the modified
Gauss law~\eqref{eq:BC_Gauss}, combined with the surface
charge-continuity equation; the term $-ik_0\chispar$ arises from the
surface divergence of the tangential polarization current.

Substituting~\eqref{eq:BC_TE_1} into~\eqref{eq:BC_TE_2} and solving
for $a_m$:
\begin{equation}\label{eq:am_TE}
a_m^{\mathrm{TE}} = -i^m\,
  \frac{(1+\chisper)\,J_m(\rho_1) - \tilde\eta_m\,J_m'(\rho_1)}
       {(1+\chisper)\,H_m^{(1)}(\rho_1) - \tilde\eta_m\,H_m^{(1)'}(\rho_1)},
  \qquad
  \tilde\eta_m = \frac{n_2 J_m(\rho_2)}{J_m'(\rho_2)} - ik_0\chispar.
\end{equation}
The standard 2D TE Mie coefficient is recovered for $\chispar =
\chisper = 0$.

\begin{remark}[Physical interpretation of the two susceptibilities]
  The two parameters enter~\eqref{eq:am_TE} in structurally distinct
  ways.  The susceptibility $\chispar$ shifts the effective interior
  impedance $\eta_m \to \tilde\eta_m$ by a purely imaginary term
  $-ik_0\chispar$: this is a reactive modification of the resonance
  condition, entirely consistent with the energy conservation result
  of Appendix~\ref{app:energy}.  The susceptibility $\chisper$, by
  contrast, renormalizes the exterior permittivity in the boundary
  condition: $(1+\chisper)$ multiplies both $J_m$ and $H_m^{(1)}$,
  in exact correspondence with the renormalized permittivity
  $\tilde\varepsilon_{\mathrm{ext}} = \varepsilon_0(1+\chisper)$
  derived in equation~\eqref{eq:eps_renorm}.  For an isotropic kernel,
  $\chispar = \chisper$, so both effects are present simultaneously.
\end{remark}

The 2D scattering efficiency is
\begin{equation}\label{eq:Q_sca}
  Q = \frac{C_{\mathrm{sca}}}{2R_0}
    = \frac{2}{k_0 \times 2R_0}
      \sum_{m=-\infty}^{+\infty}|b_m|^2
      \quad\text{(TM),}
\end{equation}
with the obvious analogue for TE (replacing $b_m$ by $a_m$).
Here $C_{\mathrm{sca}}$ is the scattering cross-section per unit
length, defined as the time-averaged scattered power per unit cylinder
length divided by the incident irradiance; $2R_0$ is the geometric
(shadow) width of the cylinder.

\subsubsection{Results}

\paragraph{Scattering efficiency vs.\ cylinder radius}

\begin{figure}[h!]
\begin{center}
\includegraphics[width=\linewidth]{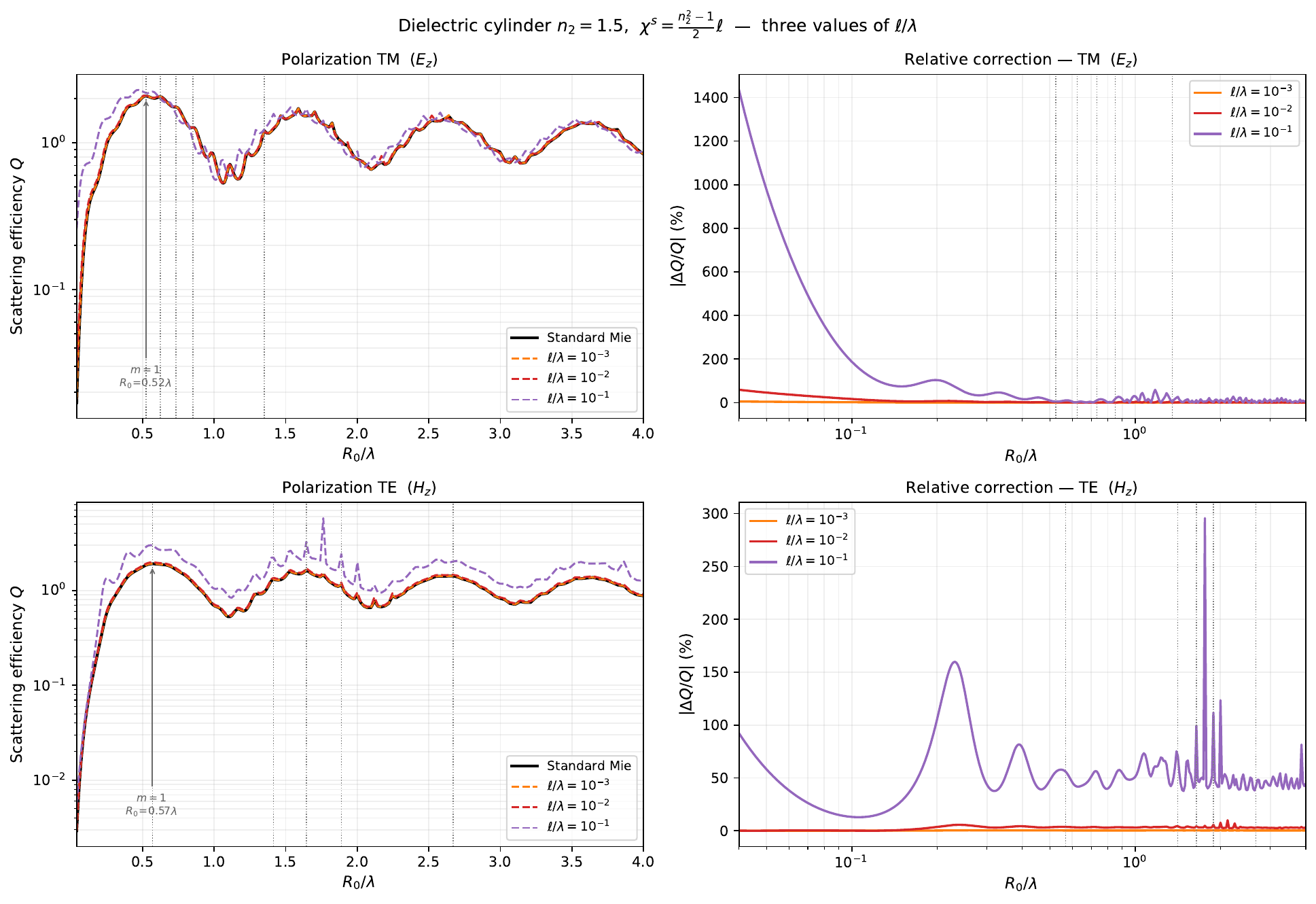}
\caption{Scattering efficiency $Q$ vs.\ $R_0/\lambda$ (left column)
and relative nonlocal correction $|\Delta Q/Q|$ (right column) for TM
(top) and TE (bottom) polarizations.  Cylinder: $n_2 = 1.5$, surface
susceptibilities from the exponential kernel~\eqref{eq:chis_consistency}.
Three values of $\ell/\lambda$ are shown: $10^{-3}$ (orange, physical
regime), $10^{-2}$ (red), $10^{-1}$ (purple).  The standard Mie result
(black) is indistinguishable from the $\ell/\lambda=10^{-3}$ curve on
the left panels; the nonlocal shift becomes clearly visible only for
$\ell/\lambda=10^{-1}$, illustrating the $\ell/\lambda$ scaling of the
correction.  Dotted vertical lines mark the Mie resonances.}
\label{fig:mie_efficiency}
\end{center}
\end{figure}

Figure~\ref{fig:mie_efficiency} shows $Q(R_0/\lambda)$ for both
polarizations and three values of $\ell/\lambda$ ($10^{-3}$, $10^{-2}$,
$10^{-1}$), together with the relative nonlocal correction
$|\Delta Q/Q| = |Q_{\mathrm{nl}} - Q_{\mathrm{std}}|/Q_{\mathrm{std}}$.
Several features are noteworthy.

\begin{enumerate}

  \item \textbf{Position of Mie resonances.}  The TM and TE resonances
    occur at different radii (TM: $R_0 \approx 0.53\lambda$; TE:
    $R_0 \approx 0.57\lambda$ for the first mode).  The nonlocal
    correction shifts these resonances by a complex amount of order
    $k_0\chispar$, modifying both their position and their quality
    factor.

  \item \textbf{Resonance enhancement of the correction.}  Away from
    resonance, the relative correction $|\Delta Q/Q|$ stays near the
    single-interface level $k_0\chispar \approx 4\times 10^{-3}$.
    At each Mie resonance, the correction is \emph{amplified} by the
    resonance quality factor: the denominator of the Mie coefficient
    becomes small, so the relative change in $b_m$ (or $a_m$) due to
    $\chispar$ is enhanced.  This amplification is the 2D Mie analogue
    of the plasmonic near-field enhancement of nonlocal effects.

  \item \textbf{Both susceptibilities matter for TE.}  The TE correction
    is generally larger than the TM correction at the same $R_0$,
    because both $\chispar$ and $\chisper$ contribute.

\end{enumerate}

\paragraph{Separate role of $\chispar$ and $\chisper$ in TE polarization.}

\begin{figure}[h!]
\begin{center}
\includegraphics[width=0.82\linewidth]{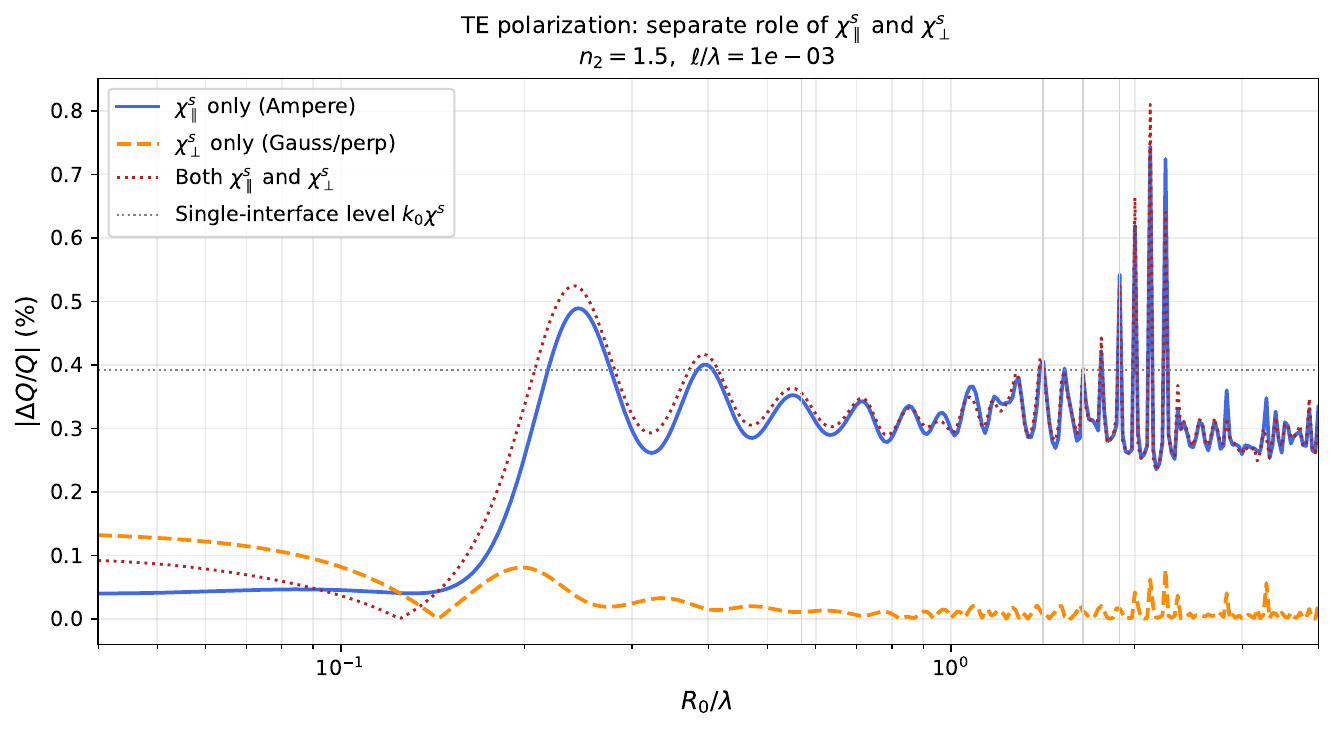}
\caption{Relative correction $|\Delta Q/Q|$ for TE polarization,
with $\chispar$ alone (blue), $\chisper$ alone (orange), and both
together (red dashed).  The two susceptibilities are set equal
($\chispar = \chisper$, isotropic kernel).  Their separate spectral
signatures are clearly distinct: $\chispar$ peaks sharply at Mie
resonances, while $\chisper$ produces a smooth, broadband baseline.}
\label{fig:mie_TE_chipar_chiperp}
\end{center}
\end{figure}

Figure~\ref{fig:mie_TE_chipar_chiperp} isolates the individual contributions
of $\chispar$ and $\chisper$ to the TE correction.  The two
susceptibilities have a clearly distinct spectral signature:
\begin{itemize}
  \item $\chispar$ (tangential, from Amp\`{e}re) dominates at
    the Mie resonances, where the complex shift of $\tilde\eta_m$
    strongly modifies the resonance condition.
  \item $\chisper$ (normal, from Gauss) acts more uniformly across
    the spectrum, producing a broadband renormalization of the exterior
    permittivity.
  \item Together they add constructively near resonance and partially
    cancel off-resonance, explaining why the combined curve in
    Figure~\ref{fig:mie_efficiency} is not simply their sum.
\end{itemize}
This separation is experimentally accessible: $\chispar$ is probed
by the tangential field (TM or TE Amp\`{e}re), while $\chisper$
requires a measurement sensitive to the normal component of $\bE$
(TM Gauss or ellipsometry at oblique incidence).

\paragraph{Near-field maps at the first Mie resonance.}

\begin{figure}[h!]
\begin{center}
\includegraphics[width=\linewidth]{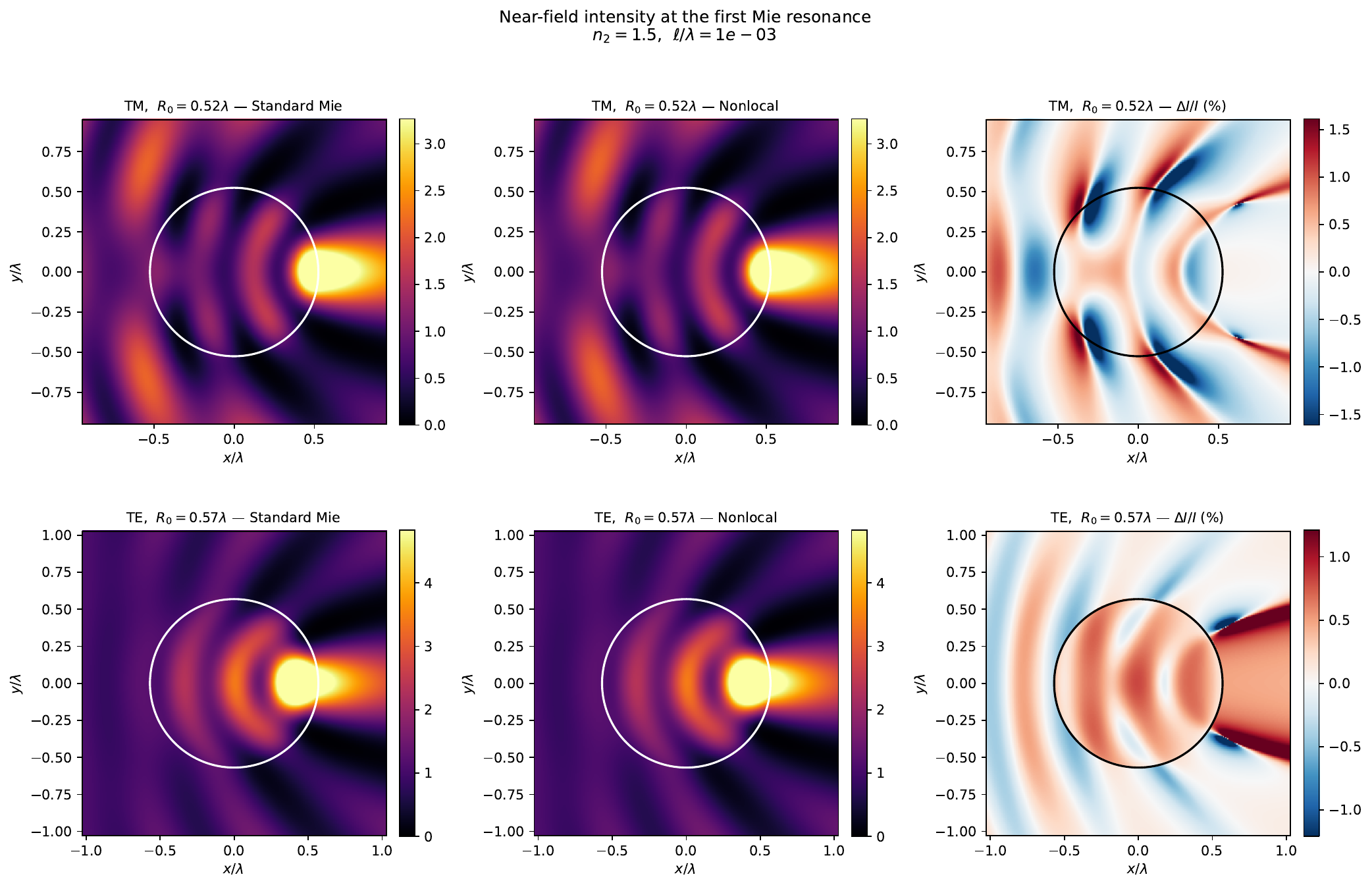}
\caption{Near-field intensity at the first Mie resonance for TM
($R_0 = 0.53\lambda$, top row) and TE ($R_0 = 0.57\lambda$, bottom row).
Left: standard Mie ($\chispar = \chisper = 0$).
Centre: nonlocal result.
Right: relative correction $\Delta I/I$ (\%).
The white (or black) circle marks the cylinder boundary.
Incident wave propagates along $+x$ (arrow).}
\label{fig:mie_nearfield}
\end{center}
\end{figure}

Figure~\ref{fig:mie_nearfield} shows the near-field intensity
$|E_z|^2$ (TM) and $|H_z|^2$ (TE) at the first resonance of each
polarization.  The relative correction $\Delta I / I$ reaches values
of order $10^{-1}$\% near the cylinder surface, concentrated in the
near-field zone $r \lesssim 2R_0$ where the evanescent contribution
of the surface susceptibility is significant.  The spatial pattern
of the correction is polarization-dependent, reflecting the distinct
roles of $\chispar$ (tangential field coupling) and $\chisper$
(normal field coupling).

\paragraph{Scaling with $R_0/\ell$.}

\begin{figure}[h!]
\begin{center}
\includegraphics[width=0.75\linewidth]{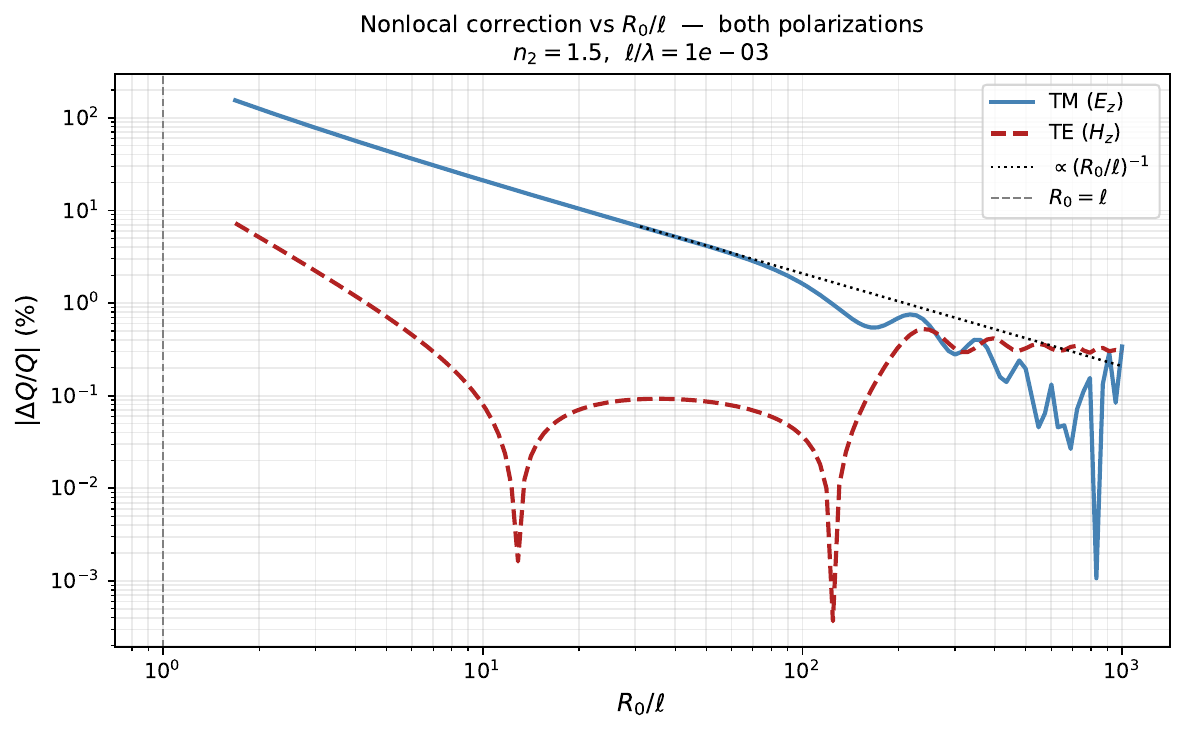}
\caption{Relative correction $|\Delta Q/Q|$ vs.\ $R_0/\ell$ for TM
(blue) and TE (red dashed) polarizations.  The black dotted line
shows the $(R_0/\ell)^{-1}$ asymptote.  The vertical dashed line
marks $R_0 = \ell$, below which the standard Mie theory is no longer
a valid approximation.  Parameters: $n_2 = 1.5$, $\ell/\lambda =
10^{-3}$.}
\label{fig:mie_R0_ell}
\end{center}
\end{figure}

Figure~\ref{fig:mie_R0_ell} displays $|\Delta Q/Q|$ as a function
of $R_0/\ell$ over four decades.  For $R_0 \gg \ell$ the correction
decays as $(R_0/\ell)^{-1}$: the surface-to-bulk ratio $\chi^s/R_0
\sim \ell/R_0 \to 0$ as the cylinder becomes large relative to the
nonlocality range.  For $R_0 \lesssim \ell$ the correction saturates
at an $O(1)$ value: the cylinder is so small that its entire volume
is within the nonlocal interaction range of the surface, and the
standard Mie theory fails entirely.

The two polarizations exhibit the same asymptotic scaling but
quantitatively different prefactors, with the TE correction
consistently larger owing to the additional $\chisper$ contribution.

\paragraph{Summary.}

Three conclusions emerge from this 2D Mie study.
\begin{enumerate}
  \item The nonlocal correction to the scattering cross section is of
    order $k_0\chispar \sim \ell/\lambda \sim 10^{-3}$ off resonance,
    but is amplified by the Mie resonance quality factor near each
    resonance frequency, making it accessible to precision scattering
    measurements.
  \item The two susceptibilities $\chispar$ and $\chisper$ can in
    principle be disentangled by comparing TM and TE measurements:
    TM is sensitive to $\chispar$ alone, while TE involves both.
  \item The nonlocal correction scales as $\ell/R_0$ for large
    particles and saturates to an $O(1)$ correction for $R_0 \lesssim
    \ell$, setting the regime in which a description based solely on
    the bulk permittivity becomes inadequate.
\end{enumerate}



\section{Conclusion}
\label{sec:conclu}

\emph{We have presented a systematic and self-contained framework for deriving
the surface electromagnetic response of a nonlocal medium from its bulk
response kernel, without any phenomenological assumption on the structure
of the transition layer.}

\paragraph{Summary of results.}

Starting from the most general tensorial nonlocal constitutive relation,
we introduced a \emph{spatial moment expansion} that naturally splits the
polarization into a bulk contribution which is identical to that of an infinite
homogeneous medium  and a boundary contribution concentrated within a
layer of thickness $\ell$ near the interface $\dOmega$.  Applying a
\emph{distributional thin-layer limit} to the boundary contribution then
converts it into a hierarchy of singular distributions supported on
$\dOmega$, whose coefficients are successive moments of the bulk kernel
$\tilde\Delta^{(1)}_{ij}$ over the exterior half-space.

For a bulk-isotropic kernel, the hierarchy simplifies considerably: the
first-order correction ($M_1$) vanishes by a parity argument, so that
curvature effects first appear at order $\ell^2$.  At leading order, the
entire interfacial physics is encoded in a single scalar, the surface
susceptibility $\chis$, given by the explicit
formula~\eqref{eq:chis3D} in terms of the zeroth-order radial moments
$\mu_0^{\parallel,\perp}$ of the kernel.  These quantities generalize the Feibelman $d$-parameters
constructively: rather than fitting parameters, they are derived
quantities determined by the bulk kernel alone.

The curvature corrections, at order $\ell^2$, are proportional to the
mean curvature $H$ and Gaussian curvature $K$ of $\dOmega$, with
coefficients $\nu^{\parallel,\perp}$ given by the second-order moments
$\mu_2^{\parallel,\perp}$.  For a plasmonic nanosphere of radius
$R_0\sim 10\,\mathrm{nm}$ with a nonlocality range $\ell\sim
0.3\,\mathrm{nm}$, these corrections are of order $1\%$, consistent with
recent experimental observations in nanophotonics
\cite{YangNature2019,Christensen2017}.

The formalism has been illustrated on four canonical geometries 
(planar, spherical, cylindrical, and ellipsoidal interfaces) and four
kernel models  (Gaussian scalar, Yukawa scalar, tensorial Lorentz), and
local limit.  In every case the calculations are fully explicit, and the
local (Fresnel) limit is recovered as a special case.  Finally, the
generalized Maxwell boundary conditions derived in Section~\ref{sec:CL}
provide a practical form of the results, directly usable in scattering
and reflection calculations.

\paragraph{Key structural observations.}

Three points deserve emphasis.
\begin{enumerate}

  \item \textbf{The kernel encodes everything.}  All surface properties :
   susceptibilities, curvature corrections, boundary conditions 
    follow from a small number of scalar moments of $\tilde\Delta^{(1)}_{ij}$.
    The microscopic complexity is fully absorbed into these moments, which
    play the role of effective parameters characterising the interface.

  \item \textbf{Surface anisotropy is a bulk property.}  For a
    scalar kernel ($\tilde\Delta_\parallel = \tilde\Delta_\perp$),
    the surface response is always isotropic ($\chispar = \chisper$),
    regardless of the interface geometry.  Anisotropy ($\chispar \neq
    \chisper$) is a signature of the tensorial structure of the bulk
    kernel, not of the curvature of the surface.

  \item \textbf{The hierarchy is universal.}  The same distributional
    structure, leading $\delta_{\dOmega}$, vanishing first-order
    correction, curvature terms at order $\ell^2$ holds for every
    geometry and every isotropic kernel.  It reflects the interplay
    between the rapid decay of nonlocal interactions and the geometric
    invariants of the surface.

\end{enumerate}

\paragraph{Outlook.}

Several natural extensions present themselves.

First, the present analysis assumed bulk centro-symmetry, which causes
the odd bulk moments to vanish.  Relaxing this assumption (e.g., for
interfaces between two different noncentrosymmetric media) introduces
additional terms in the moment expansion and modifies the boundary
conditions accordingly; the distributional framework developed here
applies without structural change.

Second, the hypothesis that the nonlocal interaction is cut off sharply
at the interface ($\mathds{1}_\Omega(\br')$ in the kernel support) can
be replaced by a smooth transition profile.  This amounts to replacing
the sharp half-space integrals by weighted integrals over a transition
zone, and does not alter the leading-order results.

Third, and most importantly, the entire formalism extends to the
\emph{nonlinear} case.  As shown in the companion
paper~\cite{ArticleNL}, the second-order surface susceptibility
$\chi^{(2),s}_{ijk}$ is derived from the bulk nonlinear kernel
$\tilde\Delta^{(2)}_{ijk}$ by exactly the same distributional procedure,
with the nonlinear kernel replaced by its linear effective kernel
obtained by integrating out one of the two field arguments.  The
hierarchy of nonlinear surface coefficients is thereby closed
recursively: $\chi^{(n)}_s$ always reduces to a $\chi^{(1)}_s$
effective problem by successive marginal integrations.  This result
provides, for the first time, a systematic derivation of the nonlinear
surface susceptibilities including $\chi^{(2)}_s$ for
second-harmonic generation at curved interfaces directly from the
bulk nonlinear kernel, without any \textit{ad hoc} surface model.

%

\appendix
\section{Isotropic rank-2 tensor functions}
\label{app:isotropy}

We give a self-contained proof that every rank-2 tensor-valued function
$F_{ij}(\bR)$ satisfying the isotropy condition
\begin{equation}\label{eq:isotropy_def}
  F_{ij}(Q\bR) = Q_{ik}\,Q_{jl}\,F_{kl}(\bR)
  \qquad \forall\, Q\in SO(3),\quad \forall\,\bR\in\R^3\setminus\{0\},
\end{equation}
has the form~\eqref{eq:iso}.

\paragraph{Step 1 — basis of available tensors.}
At a fixed $\bR\neq 0$, one seeks the symmetric rank-2 tensors that
can be formed from $\delta_{ij}$ (the only $SO(3)$-invariant rank-2
tensor) and $\bR$.  The only such tensors are $\delta_{ij}$ itself and
$R_i R_j$.  Any antisymmetric combination would require the
Levi-Civita symbol $\epsilon_{ijk}$, which is of rank~3.  Hence the
most general \emph{ansatz} consistent with isotropy is
\begin{equation}\label{eq:ansatz}
  F_{ij}(\bR) = a(R)\,\delta_{ij} + b(R)\,R_i R_j,
\end{equation}
where $a$ and $b$ can depend only on $R=|\bR|$ by the residual
$SO(2)$ symmetry around the axis $\bR$.

\paragraph{Step 2 — physical identification.}
Introduce the longitudinal projector
$P^\parallel_{ij}:=\hat R_i\hat R_j$ and the transverse projector
$P^\perp_{ij}:=\delta_{ij}-\hat R_i\hat R_j$, where
$\hat\bR := \bR/R$.  Contracting~\eqref{eq:ansatz} with each:
\begin{align*}
  F_{ij}\,\hat R_i\hat R_j &= a + b\,R^2 =: \tilde\Delta_\perp(R), \\
  F_{ij}\,e^{\perp}_i e^{\perp}_j &= a \phantom{{}+ b\,R^2}
                                    =: \tilde\Delta_\parallel(R),
\end{align*}
for any unit vector $\mathbf{e}^\perp\perp\bR$.  Solving for $a$ and $b$
and substituting back into~\eqref{eq:ansatz} yields exactly~\eqref{eq:iso}.
\qed

\begin{remark}
In Fourier space, $\tilde\Delta_\parallel(\omega,k)$ and
$\tilde\Delta_\perp(\omega,k)$ are the familiar longitudinal and
transverse dielectric functions of an isotropic
medium~\cite{Agranovich1984,LandauLifshitz1984}.  For a
centrosymmetric kernel both are even functions of $R$.
\end{remark}

\section{Derivation of the surface susceptibilities}
\label{app:chi_derivation}

We compute the zeroth-order surface moment
\begin{equation}\label{eq:S0_def_app}
  \mathcal{S}^{(0)}_{ij}
  = \int_{\Rp>0} \tilde\Delta^{(1)}_{ij}(\bR)\,\diff^3R,
\end{equation}
where $\Rp = \bR\cdot\bn > 0$ selects the exterior half-space.

\paragraph{Step 1 — insert the isotropic decomposition.}
Equation~\eqref{eq:iso} rewrites the integrand as
\begin{equation}\label{eq:iso_half}
  \tilde\Delta^{(1)}_{ij}(\bR)
  = \tilde\Delta_\parallel(R)\,\delta_{ij}
  + \bigl[\tilde\Delta_\perp(R)-\tilde\Delta_\parallel(R)\bigr]\hat R_i\hat R_j.
\end{equation}

\paragraph{Step 2 — spherical coordinates.}
Set $\bn = \hat{\mathbf{z}}$ and
$\bR = R(\sin\theta\cos\varphi,\sin\theta\sin\varphi,\cos\theta)$ with
$R>0$, $\theta\in[0,\pi/2]$, $\varphi\in[0,2\pi)$.
The relevant angular integrals over the hemisphere are:
\begin{equation}\label{eq:ang_int_app}
  \int_0^{\pi/2}\!\sin\theta\,\diff\theta = 1,\quad
  \int_0^{\pi/2}\!\sin^3\theta\,\diff\theta = \frac{2}{3},\quad
  \int_0^{\pi/2}\!\cos^2\theta\sin\theta\,\diff\theta = \frac{1}{3}.
\end{equation}

\paragraph{Step 3 — tangential component ($i=j=x$).}
Since $\hat R_x = \sin\theta\cos\varphi$, the $xx$ component of~\eqref{eq:iso_half} is
$\tilde\Delta_\parallel + (\tilde\Delta_\perp-\tilde\Delta_\parallel)\sin^2\theta\cos^2\varphi$.
Using $\int_0^{2\pi}\diff\varphi = 2\pi$ and $\int_0^{2\pi}\cos^2\varphi\,\diff\varphi = \pi$:
\begin{align}
  \mathcal{S}^{(0)}_{xx}
  &= \int_0^\infty\!\! r^2\diff r\int_0^{\pi/2}\!\!\sin\theta\,\diff\theta
     \bigl[2\pi\tilde\Delta_\parallel
           + \pi(\tilde\Delta_\perp-\tilde\Delta_\parallel)\sin^2\theta\bigr]
     \notag\\[2pt]
  &= \pi\int_0^\infty r^2\diff r\,
     \Bigl[\tilde\Delta_\parallel
           \underbrace{\int_0^{\pi/2}(2-\sin^2\theta)\sin\theta\,\diff\theta}_{=\,4/3}
          +\tilde\Delta_\perp
           \underbrace{\int_0^{\pi/2}\sin^3\theta\,\diff\theta}_{=\,2/3}\Bigr]
     \notag\\[2pt]
  &= \frac{2\pi}{3}\int_0^\infty r^2\bigl(2\tilde\Delta_\parallel(r)
     +\tilde\Delta_\perp(r)\bigr)\diff r.
     \label{eq:Sxx_app}
\end{align}

\paragraph{Step 4 — normal component ($i=j=z$).}
Since $\hat R_z = \cos\theta$, the $zz$ component of~\eqref{eq:iso_half} is
$\tilde\Delta_\parallel\sin^2\theta + \tilde\Delta_\perp\cos^2\theta$
(no $\varphi$ dependence; $\varphi$ integral gives $2\pi$):
\begin{align}
  \mathcal{S}^{(0)}_{zz}
  &= 2\pi\int_0^\infty\!\! r^2\diff r
     \Bigl[\tilde\Delta_\parallel
           \underbrace{\int_0^{\pi/2}\sin^3\theta\,\diff\theta}_{=\,2/3}
          +\tilde\Delta_\perp
           \underbrace{\int_0^{\pi/2}\cos^2\theta\sin\theta\,\diff\theta}_{=\,1/3}\Bigr]
     \notag\\[2pt]
  &= \frac{2\pi}{3}\int_0^\infty r^2\bigl(2\tilde\Delta_\parallel(r)
     +\tilde\Delta_\perp(r)\bigr)\diff r.
     \label{eq:Szz_app}
\end{align}

\paragraph{Step 5 — off-diagonal components.}
The $\varphi$ integral of $\hat R_i\hat R_j$ for $i\neq j$ involves
$\int_0^{2\pi}\cos\varphi\sin\varphi\,\diff\varphi = 0$ or
$\int_0^{2\pi}\cos\varphi\,\diff\varphi = 0$, so all off-diagonal
components vanish.

\paragraph{Result.}
Comparing~\eqref{eq:Sxx_app} and~\eqref{eq:Szz_app}, we find
$\mathcal{S}^{(0)}_{xx} = \mathcal{S}^{(0)}_{yy} = \mathcal{S}^{(0)}_{zz}$
and all off-diagonal terms are zero.  Therefore:
\begin{equation}\label{eq:S0_prop_delta}
  \mathcal{S}^{(0)}_{ij}
  = \frac{2\pi}{3}\int_0^\infty r^2
    \bigl(2\tilde\Delta_\parallel(r)+\tilde\Delta_\perp(r)\bigr)\diff r
    \,\delta_{ij}.
\end{equation}
In terms of the moments $\mu_0^\alpha = 4\pi\int_0^\infty r^2\tilde\Delta_\alpha\diff r$
(i.e., $\int_0^\infty r^2\tilde\Delta_\alpha\diff r = \mu_0^\alpha/(4\pi)$):
\begin{equation}\label{eq:S0_final_app}
  \mathcal{S}^{(0)}_{ij}
  = \frac{2\mu_0^\parallel+\mu_0^\perp}{6}\;\delta_{ij}.
\end{equation}

\begin{remark}[Proportionality to $\delta_{ij}$]
The fact that $\mathcal{S}^{(0)}_{ij}\propto\delta_{ij}$ is a theorem,
not a coincidence: integrating an SO(3)-invariant kernel over a half-space
preserves SO(2) symmetry about $\bn$, but the $\varphi$ average
additionally forces the result to be proportional to $\delta_{ij}$
(rather than $aP^\parallel_{ij}+bP^\perp_{ij}$ with $a\neq b$),
because the angular factors for the tangential and normal components
happen to be equal:
$\pi\times\tfrac{4}{3} = 2\pi\times\tfrac{2}{3}$ (tangential, Step~3) and
$\pi\times\tfrac{2}{3} = 2\pi\times\tfrac{1}{3}$ (normal, Step~4).
Consequently:
\begin{equation}\label{eq:chi_equal_app}
  \chispar = \chisper
  = \frac{2\mu_0^\parallel+\mu_0^\perp}{6}.
\end{equation}
For a centrosymmetric isotropic medium, the two surface susceptibilities
are therefore \emph{equal}.
\end{remark}

\section{Volume element in normal coordinates}
\label{app:jacobian}

Let $\dOmega$ be a smooth, closed surface in $\R^3$, and let
$\bn(\bu)$ denote the outward unit normal at the point
$\bR_0(\bu)\in\dOmega$, where $\bu=(u_1,u_2)$ are local
surface coordinates.  Any point $\br$ in a sufficiently thin
tubular neighbourhood of $\dOmega$ can be written uniquely as
\begin{equation}
  \br = \bR_0(\bu) + s\,\bn(\bu),
\end{equation}
where $s\in(-\varepsilon,\varepsilon)$ is the \emph{signed
distance} to $\dOmega$, positive in $\Omega^c$ and negative
in $\Omega$.

\paragraph{Shape operator.}
The \emph{Weingarten map} (shape operator)
$W : T_{\bR_0}\dOmega \to T_{\bR_0}\dOmega$ is defined by
$W := -\diff\bn$.
Its eigenvalues $\kappa_1,\kappa_2$ are the
\emph{principal curvatures} of $\dOmega$ at $\bR_0$, counted
positive when the surface bends toward $\Omega^c$.  The two
fundamental geometric invariants are
\begin{equation}
  H := \kappa_1 + \kappa_2
  \quad\text{(mean curvature)},
  \qquad
  K := \kappa_1\kappa_2
  \quad\text{(Gaussian curvature)}.
\end{equation}

\paragraph{Jacobian computation.}
The Jacobian of the map $(\bu,s)\mapsto\br$ is
\begin{equation}
  J(s) = \det\bigl(\mathrm{Id} + sW\bigr)
       = (1+s\kappa_1)(1+s\kappa_2)
       = 1 + s(\kappa_1+\kappa_2)
         + s^2\kappa_1\kappa_2 + O(s^3),
\end{equation}
so that the volume element reads
\begin{equation}\label{eq:jac_app}
  \diff\br
  = J(s)\,\diff S\,\diff s
  = \bigl(1 + sH + s^2 K + O(s^3)\bigr)
    \,\diff S\,\diff s,
\end{equation}
where $\diff S$ is the area element of $\dOmega$.
Equation~\eqref{eq:jac_app} is exact for the sphere and
the cylinder (for which $J(s)$ is a polynomial in $s$),
and is an asymptotic expansion in general.

\paragraph{Verification on canonical geometries.}
\begin{itemize}
  \item \emph{Plane} ($\kappa_1=\kappa_2=0$): $J(s)=1$.
  \item \emph{Sphere of radius $R_0$}
    ($\kappa_1=\kappa_2=1/R_0$):
    $J(s)=(1+s/R_0)^2 = 1+2s/R_0+s^2/R_0^2$,
    consistent with $H=2/R_0$, $K=1/R_0^2$.
  \item \emph{Cylinder of radius $R_0$}
    ($\kappa_1=1/R_0$, $\kappa_2=0$):
    $J(s)=1+s/R_0$,
    consistent with $H=1/R_0$, $K=0$.
\end{itemize}
For a proof and a complete treatment of tubular
neighbourhoods, see Gray~\cite{Gray2004}.

\section{Derivation of the generalized Fresnel coefficients}
\label{app:fresnel}

We derive the reflection and transmission coefficients
for TE polarization at a planar interface $z=0$ between a
non-dispersive medium of index $n_1$ ($z<0$) and a nonlocal
medium of bulk index $n_2$ ($z>0$), in the presence of a
surface susceptibility $\chi^s$.

\paragraph{Field ansatz.}
For TE polarization (electric field along $\hat{\mathbf{y}}$),
with the in-plane wavevector $k_x = n_1(\omega/c)\sin\theta_i
= n_2(\omega/c)\sin\theta_t$, we write
\begin{equation}
  E_y(x,z) = e^{ik_x x}
  \begin{cases}
    e^{ik_{1z}z} + r\,e^{-ik_{1z}z}, & z < 0,\\
    t\,e^{ik_{2z}z},                  & z > 0,
  \end{cases}
  \qquad k_{jz} := n_j\frac{\omega}{c}\cos\theta_j.
\end{equation}

\paragraph{Magnetic field.}
From Faraday's law $\nabla\times\mathbf{E} = i\omega\mu_0\mathbf{H}$
(with the $e^{-i\omega t}$ convention), the relevant component is
\begin{equation}
  H_x = -\frac{1}{i\omega\mu_0}\,\frac{\partial E_y}{\partial z}.
\end{equation}
Evaluating at $z = 0$:
\begin{align}
  H_x(0^-) &= -\frac{n_1\cos\theta_i}{\mu_0 c}(1-r),\\[4pt]
  H_x(0^+) &= -\frac{n_2\cos\theta_t}{\mu_0 c}\,t.
\end{align}

\paragraph{Boundary conditions.}
Two conditions apply at $z=0$.

\medskip
\noindent\textit{(i) Continuity of $E_y$ (Faraday):}
\begin{equation}\label{eq:BC1_app}
  1 + r = t.
\end{equation}

\noindent\textit{(ii) Jump of $H_x$ (generalized Amp\`{e}re,
equation~\eqref{eq:BC_plane_H}):}
\begin{equation}\label{eq:BC2_app}
  H_x(0^+) - H_x(0^-)
  = -i\omega\varepsilon_0\,\chi^s\,E_y(0)
  = -i\frac{\omega}{c}\,\frac{\chi^s}{\mu_0 c}\,(1+r).
\end{equation}
Using $1/(\mu_0 c) = \varepsilon_0 c$ and substituting the
expressions for $H_x$, condition~\eqref{eq:BC2_app} becomes
\begin{equation}\label{eq:BC2_app_expl}
  n_1\cos\theta_i(1-r)
  - n_2\cos\theta_t\,t
  = -i\frac{\omega}{c}\,\chi^s(1+r).
\end{equation}

\paragraph{Resolution.}
Substituting $t = 1+r$ from~\eqref{eq:BC1_app}
into~\eqref{eq:BC2_app_expl}:
\begin{equation}
  n_1\cos\theta_i(1-r) - n_2\cos\theta_t(1+r)
  = -i\frac{\omega}{c}\,\chi^s(1+r).
\end{equation}
Collecting terms in $r$:
\begin{equation}
  n_1\cos\theta_i - n_2\cos\theta_t
  + i\frac{\omega}{c}\,\chi^s
  = r\!\left[n_1\cos\theta_i + n_2\cos\theta_t
             - i\frac{\omega}{c}\,\chi^s\right],
\end{equation}
which gives immediately
\begin{align}
  r_{\mathrm{TE}}
  &= \frac{n_1\cos\theta_i - n_2\cos\theta_t
           + i(\omega/c)\,\chi^s}
         {n_1\cos\theta_i + n_2\cos\theta_t
           - i(\omega/c)\,\chi^s},
  \label{eq:r_TE_app}\\[8pt]
  t_{\mathrm{TE}}
  &= \frac{2n_1\cos\theta_i}
         {n_1\cos\theta_i + n_2\cos\theta_t
          - i(\omega/c)\,\chi^s}.
  \label{eq:t_TE_app}
\end{align}
For $\chi^s = 0$ one recovers the standard Fresnel
coefficients.  At normal incidence ($\cos\theta_i =
\cos\theta_t = 1$), the formula reduces to
\begin{equation}
  r_{\mathrm{TE}}\big|_{\theta=0}
  = \frac{n_1 - n_2 + i(\omega/c)\chi^s}
         {n_1 + n_2 - i(\omega/c)\chi^s},
\end{equation}
which is the relevant case for the numerical
examples of Section~\ref{sec:numerics}.

\section{Energy balance and the role of spatial non-locality}
\label{app:energy}

We examine the energy balance at a planar interface in the presence
of a surface susceptibility $\chi^s$, and address the question of
whether spatial non-locality can give rise to dissipation in analogy
with temporal non-locality.

\subsection{Temporal vs.\ spatial non-locality: a structural
            comparison}

In the temporal case, the constitutive relation reads
\begin{equation}
  P(t) = \varepsilon_0\int_{-\infty}^t \chi(t-t')\,E(t')\,\diff t',
\end{equation}
where causality imposes $\chi(\tau) = 0$ for $\tau < 0$.  In Fourier
space this becomes $\hat P(\omega) = \varepsilon_0\hat\chi(\omega)
\hat E(\omega)$, with $\hat\chi(\omega)$ complex in general.  The
Kramers--Kronig relations, which are a direct consequence of causality,
force $\mathrm{Im}(\hat\chi(\omega)) \neq 0$ whenever
$\mathrm{Re}(\hat\chi(\omega))$ is non-trivial: temporal dispersion
and absorption are structurally linked.

In the spatial case, the constitutive relation reads
\begin{equation}
  P(z) = \varepsilon_0\int_{\R}\tilde\Delta(z-z')\,E(z')\,\diff z',
\end{equation}
and there is no principle of spatial causality to constrain
$\tilde\Delta$.  If the kernel is real and even,
$\tilde\Delta(-Z) = \tilde\Delta(Z)$, its Fourier transform
$\widehat{\tilde\Delta}(k)$ is real: spatial non-locality introduces
no dissipation at the level of the bulk constitutive relation.  The
analogy with the temporal case therefore breaks down at a fundamental
level.

\subsection{Energy balance at the interface}

We now carry out the energy balance explicitly for the generalized
Fresnel problem of Appendix~\ref{app:fresnel}.  We introduce the
shorthand
\begin{equation}
  \beta_1 := n_1\cos\theta_i,
  \qquad
  \beta_2 := n_2\cos\theta_t,
  \qquad
  \gamma := \frac{\omega}{c}\,\chi^s \in \R,
\end{equation}
so that
\begin{equation}
  r_{\mathrm{TE}}
  = \frac{(\beta_1-\beta_2) + i\gamma}{(\beta_1+\beta_2) - i\gamma},
  \qquad
  t_{\mathrm{TE}}
  = \frac{2\beta_1}{(\beta_1+\beta_2) - i\gamma}.
\end{equation}

\paragraph{Reflectance and transmittance.}
The reflectance and transmittance are defined as
\begin{equation}
  R := |r_{\mathrm{TE}}|^2,
  \qquad
  T := \frac{\beta_2}{\beta_1}\,|t_{\mathrm{TE}}|^2,
\end{equation}
where the factor $\beta_2/\beta_1$ accounts for the ratio of normal
energy fluxes (Poynting vectors) on both sides of the interface.
A direct computation gives:
\begin{equation}
  R = \frac{(\beta_1-\beta_2)^2 + \gamma^2}
           {(\beta_1+\beta_2)^2 + \gamma^2},
  \qquad
  T = \frac{4\beta_1\beta_2}
           {(\beta_1+\beta_2)^2 + \gamma^2}.
\end{equation}

\paragraph{Exact conservation.}
\begin{equation}
  R + T
  = \frac{(\beta_1-\beta_2)^2 + \gamma^2 + 4\beta_1\beta_2}
         {(\beta_1+\beta_2)^2 + \gamma^2}
  = \frac{(\beta_1+\beta_2)^2 + \gamma^2}
         {(\beta_1+\beta_2)^2 + \gamma^2}
  = 1.
\end{equation}
Energy is \emph{exactly} conserved for any real value of $\chi^s$,
regardless of the angle of incidence or the indices.  The surface
susceptibility, however large, does not introduce any absorption.

\subsection{Physical interpretation: reactive vs.\ dissipative
            response}

The appearance of $i\gamma$ in the numerator and denominator of
$r_{\mathrm{TE}}$ might suggest dissipation, since it shifts the
pole structure of the Fresnel coefficient.  The energy balance above
shows that this is not the case: the surface term acts as a purely
\emph{reactive} element, analogous to a lossless capacitor or
inductor in circuit theory.

This can be understood directly from the boundary
condition~\eqref{eq:BC_plane_H}:
\begin{equation}
  \bn\times\jump{\bH} = -i\omega\varepsilon_0\,\chi^s\,\bE^-_\parallel.
\end{equation}
For real $\chi^s$ and harmonic fields ($e^{-i\omega t}$), the surface
current $\mathbf{J}^s := \bn\times\jump{\bH}$ is in quadrature with
the field $\bE^-_\parallel$: the time-averaged power dissipated in
the surface,
\begin{equation}
  \mathcal{P}^s
  = \frac{1}{2}\,\mathrm{Re}
    \bigl(\mathbf{J}^s\cdot\bE^{-*}_\parallel\bigr)
  = \frac{\omega\varepsilon_0\chi^s}{2}
    \,\mathrm{Re}\bigl(i|\bE^-_\parallel|^2\bigr)
  = 0,
\end{equation}
is identically zero.  The surface term redistributes energy between
the reflected and transmitted waves without absorbing any.

\subsection{Effect on the phase of the reflection coefficient}

Although $|r_{\mathrm{TE}}|^2$ is modified at order $\gamma^2$ only
(since the first-order term in $\gamma$ is purely imaginary), the
\emph{phase} of $r_{\mathrm{TE}}$ is sensitive to $\gamma$ at first
order.  Writing $r_{\mathrm{TE}} = |r_{\mathrm{TE}}|\,e^{i\phi}$:
\begin{equation}
  \phi
  = \arg\!\bigl[(\beta_1-\beta_2)+i\gamma\bigr]
    - \arg\!\bigl[(\beta_1+\beta_2)-i\gamma\bigr]
  = \arctan\!\frac{\gamma}{\beta_1-\beta_2}
    + \arctan\!\frac{\gamma}{\beta_1+\beta_2},
\end{equation}
and for $\gamma \ll \beta_1, \beta_2$:
\begin{equation}
  \phi \approx \phi_0
  + \gamma\!\left(\frac{1}{\beta_1-\beta_2}
                 +\frac{1}{\beta_1+\beta_2}\right)
  + O(\gamma^2)
  = \phi_0
  + \frac{2\beta_1\gamma}{\beta_1^2 - \beta_2^2}
  + O(\gamma^2),
\end{equation}
where $\phi_0 = \arg(r_0) \in \{0,\pi\}$ for real indices.
The phase shift is linear in $\gamma \propto \chi^s\ell/\lambda$
and is the primary observable signature of the surface
susceptibility in the transparent case.  This is why ellipsometry,
which measures the ratio $r_p/r_s$ and is directly sensitive to
phase differences, is the technique of choice for detecting
non-local surface effects at transparent interfaces
\cite{YangNature2019}.

\subsection{When does spatial non-locality produce dissipation?}

The above analysis assumed $\chi^s \in \R$.  We now identify the
physical situations in which $\chi^s$ becomes complex and genuine
dissipation arises.

\paragraph{(i) Frequency-dependent kernel.}
At finite frequency $\omega$, the bulk kernel $\tilde\Delta$ depends
on $\omega$ through the electronic response of the medium.
Integrating out the electronic degrees of freedom generically yields
a complex $\tilde\Delta(\cdot,\omega)$, and hence a complex
$\chi^s(\omega)$.  In this case,
\begin{equation}
  \mathrm{Im}(\chi^s(\omega)) \neq 0,
\end{equation}
and the Kramers--Kronig relations now apply to $\chi^s(\omega)$,
linking spatial dispersion and surface absorption.  This is precisely
the mechanism responsible for Landau damping at metal surfaces~\cite{Landau1946}.

\paragraph{(ii) Non-centrosymmetric kernel.}
If the bulk kernel is real but not even,
$\tilde\Delta(-Z) \neq \tilde\Delta(Z)$, its Fourier transform
acquires a non-zero imaginary part:
\begin{equation}
  \widehat{\tilde\Delta}(k)
  = \int_\R \tilde\Delta(Z)\,e^{ikZ}\,\diff Z
  = \widehat{\tilde\Delta}_{\mathrm{even}}(k)
    + i\,\widehat{\tilde\Delta}_{\mathrm{odd}}(k).
\end{equation}
The surface susceptibility $\chi^s$ then inherits an imaginary part
from the odd component of the kernel.  The energy balance becomes:
\begin{equation}
  R + T = 1 - \mathcal{A},
  \qquad
  \mathcal{A}
  = \frac{4\beta_1\,(\omega/c)\,\mathrm{Im}(\chi^s)}
         {(\beta_1+\beta_2)^2 + |(\omega/c)\chi^s|^2}
  \geq 0,
\end{equation}
where $\mathcal{A} \geq 0$ is the fraction of incident energy
absorbed in the surface layer.  The inequality $\mathcal{A} \geq 0$
requires $\mathrm{Im}(\chi^s) \geq 0$, which is the surface analogue
of the bulk condition $\mathrm{Im}(\hat\chi(\omega)) \geq 0$
(passivity of the medium).

\paragraph{(iii) Propagating surface modes.}
Even for a real $\chi^s$, the modified Fresnel denominator
$(\beta_1+\beta_2) - i\gamma$ can have a zero for complex angles of
incidence, corresponding to a surface polariton whose radiation
damping is modified by the non-local surface term.  This is not
dissipation in the thermodynamic sense, but a redistribution of
energy into a guided surface mode.

\begin{remark}[Summary]
  Spatial non-locality, for a real and even kernel, is a purely
  reactive phenomenon: it shifts the phase of the reflected wave and
  modifies the balance between $R$ and $T$ only at order
  $(\ell/\lambda)^2$, but conserves energy exactly.  Dissipation
  arises only when $\mathrm{Im}(\chi^s)\neq 0$, which requires
  either a frequency-dependent (absorbing) bulk kernel or a
  non-centrosymmetric kernel.  In both cases the condition
  $\mathrm{Im}(\chi^s)\geq 0$ is the surface expression of the
  passivity of the medium, in direct analogy with the bulk condition
  $\mathrm{Im}(\hat\varepsilon(\omega))\geq 0$.
\end{remark}

\section{Surface susceptibility as a thin-layer limit}
\label{app:thin_layer}

We show how the surface susceptibility $\chi^s$ arises naturally from
a thin homogeneous layer of thickness $L$ and local susceptibility
$\chi_{\mathrm{lay}}$, in the limit $L\to 0$ with the product
$\chi_{\mathrm{lay}} L$ kept fixed.

\subsection{Setup}

\begin{figure}[h!]
\begin{center}
\begin{tikzpicture}[>=stealth, thick]

\pgfmathsetmacro{\Ltot}{9}
\pgfmathsetmacro{\Ht}{4.5}
\pgfmathsetmacro{\xL}{3.0}
\pgfmathsetmacro{\xR}{4.2}
\pgfmathsetmacro{\xMidL}{0.5*\xL}
\pgfmathsetmacro{\xMid}{0.5*(\xL+\xR)}
\pgfmathsetmacro{\xMidR}{0.5*(\xR+\Ltot)}
\pgfmathsetmacro{\yinc}{0.38*\Ht}
\pgfmathsetmacro{\yref}{0.55*\Ht}
\pgfmathsetmacro{\ylay}{0.46*\Ht}
\pgfmathsetmacro{\ytrans}{0.46*\Ht}
\pgfmathsetmacro{\yaxis}{0.5*\Ht}
\pgfmathsetmacro{\ytoplab}{0.88*\Ht}
\pgfmathsetmacro{\xLm}{(\xL-0.15)}
\pgfmathsetmacro{\xLp}{(\xL+0.10)}
\pgfmathsetmacro{\xRm}{(\xR-0.10)}
\pgfmathsetmacro{\xRp}{(\xR+0.15)}
\pgfmathsetmacro{\xEnd}{(\Ltot-0.3)}
\pgfmathsetmacro{\xLtot}{(\Ltot+0.3)}
\pgfmathsetmacro{\yannA}{0.18*\Ht}
\pgfmathsetmacro{\yannB}{0.08*\Ht}
\pgfmathsetmacro{\xannB}{(\xR+1.4)}

\fill[blue!6]  (0,0)   rectangle (\xL,\Ht);
\fill[gray!20] (\xL,0) rectangle (\xR,\Ht);
\fill[blue!6]  (\xR,0) rectangle (\Ltot,\Ht);

\fill[pattern=north east lines, pattern color=gray!50]
    (\xL,0) rectangle (\xR,\Ht);

\draw[gray!70, dashed, thick] (\xL,0) -- (\xL,\Ht);
\draw[gray!70, dashed, thick] (\xR,0) -- (\xR,\Ht);

\node[align=center] at (\xMidL,\ytoplab)
    {\small Medium 1 \\ \small index $n_1$};
\node[align=center] at (\xMid,\ytoplab)
    {\small Layer \\ \small $\chi_{\mathrm{lay}},\; L$};
\node[align=center] at (\xMidR,\ytoplab)
    {\small Medium 2 \\ \small index $n_2$};

\draw[gray!60] (\xL,-0.1) -- (\xL,-0.6);
\draw[gray!60] (\xR,-0.1) -- (\xR,-0.6);
\draw[<->] (\xL,-0.45) -- (\xR,-0.45)
    node[midway, below] {\small $L$};
\node[below] at (\xL,-0.65) {\small $z=0$};
\node[below] at (\xR,-0.65) {\small $z=L$};

\draw[->] (-0.3,\yaxis) -- (\xLtot,\yaxis) node[right] {$z$};

\draw[->, color=blue!70!black, line width=1.4pt]
    (0.3,\yinc) -- (\xLm,\yinc);
\node[above, color=blue!70!black] at (\xMidL,\yinc)
    {\small $E^{\mathrm{inc}}$};

\draw[->, color=red!70!black, line width=1.4pt]
    (\xLm,\yref) -- (0.3,\yref);
\node[above, color=red!70!black] at (\xMidL,\yref)
    {\small $r\,E^{\mathrm{inc}}$};

\draw[->, color=gray!60!black, line width=1.2pt]
    (\xLp,\ylay) -- (\xRm,\ylay);
\node[above, color=gray!60!black] at (\xMid,\ylay)
    {\small $E^{\mathrm{lay}}$};

\draw[->, color=teal!70!black, line width=1.4pt]
    (\xRp,\ytrans) -- (\xEnd,\ytrans);
\node[above, color=teal!70!black] at (\xMidR,\ytrans)
    {\small $t\,E^{\mathrm{inc}}$};

\draw[<-, gray!50]
    (\xMid,\yannA) to[out=-90, in=180] (\xannB,\yannB);
\node[right, gray!60] at (\xannB,\yannB)
    {\small $L\to 0,\;\chi_{\mathrm{lay}} L = \chi^s$ fixed};

\end{tikzpicture}
\caption{Geometry of the thin-layer model. A homogeneous layer of
thickness $L$ and local susceptibility $\chi_{\mathrm{lay}}$ is
sandwiched between medium~1 (index $n_1$, $z<0$) and medium~2
(index $n_2$, $z>L$). In the thin-layer limit $L\to 0$ with
$\chi_{\mathrm{lay}}L = \chi^s$ fixed, the layer collapses onto
the interface $z=0$ and its integrated response is entirely
encoded in the surface susceptibility $\chi^s$.}
\label{fig:thin_layer}
\end{center}
\end{figure}
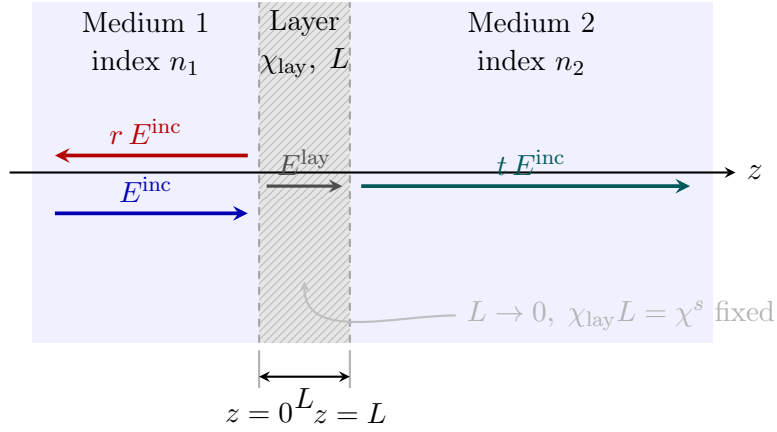

We consider three regions (see Fig.~\ref{fig:thin_layer}):
\begin{itemize}
  \item medium 1 ($z < 0$), index $n_1$, susceptibility
    $\chi_1 = n_1^2 - 1$;
  \item a homogeneous layer ($0 < z < L$), local susceptibility
    $\chi_{\mathrm{lay}}$, index $n_{\mathrm{lay}}^2 = 1 +
    \chi_{\mathrm{lay}}$;
  \item medium 2 ($z > L$), index $n_2$, susceptibility
    $\chi_2 = n_2^2 - 1$.
\end{itemize}
For TE polarization at normal incidence, the electric field satisfies
the Helmholtz equation in each region, and the standard continuity
conditions (continuity of $E_y$ and $\partial_z E_y$) apply at $z=0$
and $z=L$.

\subsection{Transfer matrix of the layer}

The propagation matrix of the layer of thickness $L$ and wavenumber
$k_{\mathrm{lay}} = n_{\mathrm{lay}}\,\omega/c$ is
\begin{equation}
  \mathbf{P}(L)
  =
  \begin{pmatrix}
    \cos(k_{\mathrm{lay}} L)
    & \dfrac{-i}{n_{\mathrm{lay}}}\sin(k_{\mathrm{lay}} L)
    \\[8pt]
    -i n_{\mathrm{lay}}\sin(k_{\mathrm{lay}} L)
    & \cos(k_{\mathrm{lay}} L)
  \end{pmatrix}.
\end{equation}
We expand for $k_{\mathrm{lay}} L \ll 1$:
\begin{equation}\label{eq:P_expand}
  \mathbf{P}(L)
  =
  \begin{pmatrix}
    1 & 0 \\ 0 & 1
  \end{pmatrix}
  +
  k_{\mathrm{lay}} L
  \begin{pmatrix}
    0 & -i/n_{\mathrm{lay}} \\
    -i n_{\mathrm{lay}} & 0
  \end{pmatrix}
  + O\!\left((k_{\mathrm{lay}} L)^2\right).
\end{equation}

\subsection{Thin-layer limit}

As $L \to 0$ with $\chi_{\mathrm{lay}} L = \chi^s$ fixed, the
wavenumber $k_{\mathrm{lay}} = (\omega/c)\sqrt{1+\chi_{\mathrm{lay}}}$
diverges as $(\omega/c)\sqrt{\chi_{\mathrm{lay}}}$, but the relevant
product remains finite:
\begin{equation}
  k_{\mathrm{lay}} L \cdot n_{\mathrm{lay}}
  = \frac{\omega}{c}\,n_{\mathrm{lay}}^2\,L
  = \frac{\omega}{c}(1+\chi_{\mathrm{lay}})L
  \;\xrightarrow{L\to 0}\;
  \frac{\omega}{c}\,\chi^s,
\end{equation}
where we used $\chi_{\mathrm{lay}} L \to \chi^s$ and $L \to 0$.
The off-diagonal terms of~\eqref{eq:P_expand} then give, in the
limit:
\begin{equation}\label{eq:P_limit}
  \mathbf{P}(L)
  \;\xrightarrow{L\to 0}\;
  \begin{pmatrix}
    1 & 0 \\
    -i\,(\omega/c)\,\chi^s & 1
  \end{pmatrix}
  =: \mathbf{I}(\chi^s).
\end{equation}
This interface matrix encodes the surface susceptibility correction to the standard transfer-matrix formalism.

\subsection{Effective boundary conditions}

Applying the transfer matrix $\mathbf{I}(\chi^s)$ at $z = 0$,
one obtains the jump conditions between medium 1 and medium 2:
\begin{align}
  E_y^+ &= E_y^-,
  \label{eq:BC_E_layer}\\[6pt]
  \frac{\partial_z E_y^+}{i\omega\mu_0}
  - \frac{\partial_z E_y^-}{i\omega\mu_0}
  &= -i\omega\varepsilon_0\,\chi^s\,E_y^-,
  \label{eq:BC_H_layer}
\end{align}
i.e., in terms of the magnetic field $H_x = -\partial_z E_y /
(i\omega\mu_0)$:
\begin{equation}
  \jump{H_x} = -i\omega\varepsilon_0\,\chi^s\,E_y,
\end{equation}
which is identical to the boundary condition~\eqref{eq:BC_plane_H}
derived from the nonlocal framework. The surface susceptibility
$\chi^s$ is therefore the effective parameter encoding the
integrated response of the layer in the limit of vanishing
thickness.

\subsection{Physical content of the limit}

The thin-layer limit $L \to 0$, $\chi_{\mathrm{lay}} L = \chi^s$
fixed has a precise physical meaning.  Three remarks are in order.

\paragraph{Consistency with the moment expansion.}
In the nonlocal framework, $\chi^s$ is given by the first moment
of the kernel (equation~\eqref{eq:A0_final}):
\begin{equation}
  \chi^s = \int_0^{+\infty} Z\,\tilde\Delta(Z)\,\diff Z.
\end{equation}
For the exponential kernel $\tilde\Delta(Z) = (A/\ell)\,e^{-Z/\ell}$,
this gives $\chi^s = A\ell$.  The thin-layer model reproduces
this result with the identification
\begin{equation}
  \chi_{\mathrm{lay}} L = A\ell,
\end{equation}
provided the layer thickness $L$ is identified with the nonlocality
range $\ell$ and the layer susceptibility with the kernel amplitude
$\chi_{\mathrm{lay}} = A$.

\paragraph{Energy conservation.}
Since the matrix $\mathbf{I}(\chi^s)$ is unimodular
($\det \mathbf{I} = 1$) for real $\chi^s$, the thin-layer limit
preserves the energy flux: $R + T = 1$ exactly, consistently
with Appendix~\ref{app:energy}.

\paragraph{Breakdown of the limit.}
The expansion~\eqref{eq:P_expand} requires $k_{\mathrm{lay}} L \ll 1$,
i.e., $L \ll \lambda/n_{\mathrm{lay}}$.  For a metallic layer with
$n_{\mathrm{lay}} \sim 10$ at optical frequencies, this restricts the
validity of the thin-layer approximation to $L \lesssim 5$\,nm,
precisely the range in which the surface susceptibility description
is expected to be adequate, and beyond which higher-order terms in
the moment expansion (proportional to $\delta'_{\dOmega}$,
$\delta''_{\dOmega}$, etc.) become necessary.

\begin{remark}[Connection with the Drude--Smith model]
  For a Drude metal layer~\cite{Drude1900,Smith2001} of plasma frequency $\omega_p$ and
  thickness $L \ll c/\omega_p$, the layer susceptibility at
  frequency $\omega$ is $\chi_{\mathrm{lay}}(\omega) \approx
  -\omega_p^2/\omega^2$.  The thin-layer limit then gives
  \begin{equation}
    \chi^s(\omega) = -\frac{\omega_p^2 L}{\omega^2},
  \end{equation}
  which is purely real (no dissipation in this limit) and
  negative, corresponding to a capacitive surface response.
  The imaginary part of $\chi^s(\omega)$, responsible for
  surface absorption, appears only when damping is included
  in the Drude model, consistently with the discussion of
  Appendix~\ref{app:energy}.
\end{remark}
\printbibliography

\end{document}